\documentclass{aa}  

\usepackage{txfonts}
\bibpunct{(}{)}{;}{a}{}{,}
\graphicspath{ {paper_figs/} }

\newcommand{\Ms}{{M_\odot}}

\newcommand{\cc}{{{\rm cm}^{-3}}}

\begin{document}

   \title{Stellar mass spectrum within massive collapsing clumps \\ III. Effects of temperature and magnetic field}

   \subtitle{}

   \titlerunning{Effects of temperature and magnetic field on the IMF}

   \author{
          Yueh-Ning Lee \inst{1,2,3}
          \and
          Patrick Hennebelle\inst{2,3,4}%\fnmsep\thanks{}
          }
   \institute{Institut de Physique du Globe de Paris, Sorbonne Paris Cit\'e, Universit\'e Paris Diderot, UMR 7154 CNRS, F-75005 Paris, France\\
              \email{ynlee@ipgp.fr}
         \and
                Universit\'{e} Paris Diderot, AIM, Sorbonne Paris Cit\'{e}, CEA, CNRS, F-91191 Gif-sur-Yvette, France
         \and
                IRFU, CEA, Universit\'{e} Paris-Saclay, F-91191 Gif-sur-Yvette, France
         \and
             LERMA (UMR CNRS 8112), Ecole Normale Sup\'{e}rieure, 75231 Paris Cedex, France\\
             \email{patrick.hennebelle@lra.ens.fr }
%             %\thanks{}
             }

  \date{Accepted 13/12/2018}

% \abstract{}{}{}{}{} 
% 5 {} token are mandatory
 
  \abstract
   {The stellar mass spectrum is an important property of the stellar cluster and a fundamental quantity to understand our Universe. The fragmentation of diffuse molecular cloud into stars is subject to physical processes such as gravity, turbulence, thermal pressure, and  magnetic field.}
   {The final mass of a star is believed to be a combined outcome of a virially unstable reservoir and subsequent accretion. We aim to clarify the roles of different supporting energies, notably the thermal pressure and the magnetic field, in determining the stellar mass. }
   {Following previous studies by \citet{Lee18a, Lee18b}, we perform a series of numerical experiments of stellar cluster formation inside an isolated molecular clump. By changing the effective equation of state (EOS) of the diffuse gas (that is to say gas whose density is below the critical density at which dust becomes opaque to its radiation) and the strength of the magnetic field, we investigate whether any characteristic mass  is  introduced into the fragmentation processes.}
   {The EOS of the diffuse gas, including the bulk temperature and the polytropic index, does not affect significantly the shape of the stellar mass spectrum. The presence of magnetic field slightly modifies the shape of the mass spectrum only when extreme values are applied. }
   {This study confirms that the peak of the IMF is primarily determined by the adiabatic high-density end of the EOS that mimics the radiation inside the high-density gas. Furthermore, the shape of the mass spectrum is mostly sensitive to the density PDF, and the magnetic field has likely only a secondary role. In particular, we stress that the Jeans mass at the mean cloud density and at the critical  density are not responsible of setting the peak.}


   \maketitle

%________________________________________________________________________________
%\tableofcontents
%------------------INTRODUCTION------------------------
\section{Introduction}
Deciphering the origin of the initial mass function (IMF), which describes the mass spectrum of stars at the moment of their birth  \citep[e.g.,][]{Kroupa01,Chabrier03,Bastian10,Offner14}, 
is crucial for our understanding of the Universe since stars are the driving engine of the evolution of large structures. 
The stellar mass spectrum is the outcome of molecular cloud fragmentation, 
which is a process subject to the gravity and turbulence, 
as well as other physical factors, 
such as the thermal pressure, radiation, magnetic field, and stellar feedback. 
Observations over decades suggest a relatively universal form of the IMF irrespectively of the star-forming conditions, except for some recent works \citep{Cappellari12,Hosek18} that suggest possible variations. 
In particular, the IMF exhibits a characteristic peak at $\sim 0.2-0.3~\Ms$, with small variations observed exclusively in extreme environments.  
The high-mass end is usually described with a powerlaw such that $dN/d\log M \propto M^\Gamma$, 
with $\Gamma = -1.35$ typically \citep{Salpeter55}.

From the most classical point of view, 
the fragmentation of a medium is characterized by the Jeans instability criteria, 
which is based on the density and temperature of the molecular cloud, 
with other physics playing probably secondary roles. 
Simulations including various physics have been performed by many authors to explain the characteristic mass of the IMF \citep[e.g.,][]{Bate03, Bonnell03, Offner08,Girichidis11,Krumholz11,BallesterosParedes15}, 
with theoretical models being developed in parallel \citep[e.g.][]{Padoan97, Inutsuka01,HC08, Hopkins12}.
In this work, we focus particularly on the link between the characteristic value of the stellar mass spectrum from numerical simulations of molecular cloud collapse and the Jeans mass and magnetic 
Jeans mass of the initial cloud.

In previous studies, we have examined the effects of initial cloud density and turbulent support, 
as well as numerical resolutions \citep[][hereafter Paper I]{Lee18a}, 
and showed that the powerlaw slope at the high-mass end of the stellar mass spectrum depends on the relative importance of the turbulent, thermal, and gravitational energy. 
When the turbulent energy dominates over thermal energy, the fragmentation produces a powerlaw mass distribution $dN/d\log M \propto M^{-0.75}$, 
while this distribution becomes flat otherwise. In particular by varying the mean density of the cloud by several
orders of magnitude, 
we show in paper I that the mean Jeans 
mass within the cloud does not affect the peak position.
We  proposed a mechanism to explain the universality of the IMF peak \citep[][hereafter Paper II]{Lee18b}, 
which is based on the mass of the first hydrostatic core (also called the first Larson core).
The first Larson core defines the minimum mass required to trigger the second collapse that forms the protostar, 
and the tidal field around it prevents nearby fragmentation and thus leads to further mass accretion, 
yielding a final stellar mass that is about ten times that of the first Larson core. 
A more detailed statistical model was developed to account for this factor 10 (Hennebelle et al. submitted to A\&A).

In this study, we consider other factors that may have an impact on the cloud fragmentation, 
notably the temperature and the magnetic field.  
The thermal and magnetic energy act against self-gravity and could provide a characteristic scale for the fragmentation. 
The goal of this current study is to determine whether such characteristic scale does exist and is reflected in the mass spectra from cluster formation simulations.  
In contrary to what is intuitively expected, 
the peak of the stellar mass spectrum resulting from this study does not depend on the Jeans mass
taken either at the mean cloud density or at the critical density. 
 The characteristic mass being due to solely to the first hydrostatic core, i.e. the 
high density part of the effective equation of state (EOS)
 and surrounding collapsing envelope  (Paper II) is confirmed.

The plan of this paper is as follows:
In Sect. 2, the simulation setups are described in details. 
The EOS of the gas and the magnetic field strength are varied. 
Sink particles are used to trace the formation and accretion of stars.
In Sect. 3, the stellar mass spectra resulting from the simulations are presented and commented qualitatively. 
In Sect. 4, the effects of the varied parameters are discussed and interpreted more quantitatively. 
Finally, Sect. 5 concludes the paper.

\section{Numerical simulations}

\subsection{The temperature}
\begin{table*}[]
\caption{Parameters for simulations with varied EOS. The label of the run, the bulk temperature, the cloud central density, the adiabatic critical density, the sink density threshold, the polytropic indices, the maximal refinement level, and the physical resolution are listed }
\label{table_T_params}
\centering
\begin{tabular}{l c c c c c c c}
\hline\hline
Label   & $T_0$(K) & $n _0 (~\cc)$ & $n _{\rm ad} (10^{10} \cc)$ & $n _\mathrm{sink} (10^{10} \cc)$ &  $\gamma_0/\gamma$  &  $l_{\rm max}$    & resolution (AU) \vspace{.5mm}\\
\hline
T10M0+   & $10$& $6.0 \times 10^7$ & $1$  & $1$ &  1/1.66  & 15 & 2   \\%Lowmass_compact_relax_l8_2_res                                                      
T40    & $40$ & $2.8 \times 10^9$&  $8.17$  & $30$ &  1/1.66  & 12 & 4   \\%Lowmass_compact_relax_l8_40K                                             
T100   & $100$ & $6.0 \times 10^{10}$ & $32.7$  & $30$ & 1/1.66  & 11 & 3   \\%Lowmass_compact_relax_l8_100K      
\hline
T10G12 &  $10$& $6.0 \times 10^7$ & $1$  & $1$ &  1.2/1.66  & 14 & 4   \\%C_p02 
T10G07 &  $10$& $6.0 \times 10^7$ & $1$  & $1$ &  0.7/1.66  & 14 & 4   \\%C_n03                                       
\hline
\end{tabular}
\end{table*}

The thermal energy is one of the major support against the gravity at prestellar core scales. 
We designed a series of simulations to investigated the effects of the thermal behaviors at low densities on the gas fragmentation outcome. 
A barotropic EOS prescription is used for the temperature such that
\begin{align}\label{eq_EOS2}
T(n) = T_0 \left[ \left({n \over n_{\rm ad}}\right)^{\gamma_0-1}+ \left({n \over n_{\rm ad}}\right)^{\gamma-1} \right],
\end{align}
where $T_0$ is the temperature of the diffuse bulk gas, $n$ is the gas number density, 
$n_{\rm ad}$ is the critical density at which the gas turns adiabatic, 
$\gamma_0$ and $\gamma$ are the polytropic indices at for the diffuse and dense gas, respectively. 
Two effects are considered: 
first $T_0$ is varied, 
and secondly a shallow polytropic index $\gamma_0$ other than 1 at low densities is considered. 
As  suggested by previous studies that isothermal gas has no lower mass limit for fragmentation 
\citep[Paper I;][]{Guszejnov18} and 
an adiabatic EOS (with $\gamma \ge 4/3$) is necessary to yield a characteristic mass. 
Here we would like to verify that the low-density part of the EOS, that 
is to say below the critical density at which the gas becomes adiabatic, plays indeed no major role.

The classical expression for the Jeans mass is given by 
\begin{align}
M_{\rm J} &= {\pi \over 6} {c_{\rm s}^3 \over \sqrt{G^3 \rho}} = {\pi \over 6} \left({k_{\rm B}\over \mu m_{\rm p}G}\right)^{3/2}\rho^{-1/2} T^{3/2}  \nonumber \\
&=  5.1 \times 10^{-4} \Ms \left({n\over 10^{10}~ \cc}\right)^{-1/2} \left({T \over 10 ~{\rm K}}\right)^{3/2} , 
\label{Mjeans}
\end{align}
where the Jeans mass is defined as the mass contained in a sphere of diameter that is equal to the Jeans length, 
$\rho = n \mu m_{\rm p}$ is the density, $T$ is the temperature, 
$k_{\rm B}$ is the Boltzmann constant, $\mu=2.33$ is the mean molecular weight, $m_{\rm p}$ is the proton mass, 
and $G$ is the gravitational constant. 
For the EOS as suggested by Eq. (\ref{eq_EOS2}) with $\gamma_0 = 1$, the Jeans mass decreases as $n^{-0.5}$ at low densities and increases as $n^{0.5}$ above the critical density of adiabatic heating, $n_{\rm ad}$. 
The Jeans mass around $n_{\rm ad}$ has been traditionally considered to set the minimum mass for 
fragmentation and the question arises as to whether it may also set the peak of the IMF. 
On the contrary the argument proposed in paper II only invokes the mass of the first hydrostatic core and 
is independent of Eq.~(\ref{Mjeans}).

With a view to test this argument for characteristic fragmentation mass based on the Jeans mass, 
we vary the EOS behavior in the low density regime in a series of simulations. 
The run C from paper I is taken as a reference. 
The initial cloud has $M=1000~\Ms$ and is spherical with a density profile $\rho(r) = \rho_0/\left[1+\left(r/r_0\right)^2\right]$, where $r$ is the distance to the cloud center, and
 $\rho_0$ and $r_0$ are the density and size of the central plateau, respectively. 
The density contrast between the cloud center and edge is set to ten, and the radius consequently is $3r_0$.
The simulation box size is twice the cloud diameter and the rest of the space is patched with a uniform diffuse medium of $\rho_0/100$. 
The turbulence is initially seeded with a Kolmogorov spectrum with random phases and no turbulence driving is employed. 
The amplitude of the turbulence is scaled to match the assigned Mach number.

\subsubsection{Bulk temperature of the diffuse gas}
To investigate the effect of cloud overall temperature on the characteristic fragmentation mass, 
we performed a series of simulations with varied bulk temperature, $T_0$, while keeping $\gamma_0=1$ in Eq. (\ref{eq_EOS2}). 
This implies that the gas is isothermal at low densities, and heats up adiabatically due to dust opacity at higher densities with $\gamma = 1.66$, 
which is a reasonable approximation of the interstellar medium up to the temperature of hydrogen molecule dissociation. 
We recall that in Paper II, the high-density end of the EOS was varied. 
More precisely, we studied different values of $n_{\rm ad}$ and $\gamma$ and found a dependence of the mass spectra peak mass on the mass of the first hydrostatic core. 

The simulations in Papers I and II had $T_0$ fixed to 10 K. 
In this study, values of 40, and 100 K are used instead while keeping all the other parameters identical. 
The critical density $n_{\rm ad} = 10^{10} ~\cc$ in the canonical run. 
For the other runs, $n_{\rm ad}$ is chosen such that the upper end of the EOS coincides with the canonical one
(see Fig.\ref{fig_EOS_all}), 
giving 
\begin{eqnarray}
n_{\rm ad}  = 10^{10}~ \cc \left( T_0/10~ {\rm K}\right)^{1/(\gamma-1)}.
\end{eqnarray}
This choice ensures that the mass of the first hydrostatic core is unchanged while the Jeans mass 
at the critical density varies (see discussion in Sect.~\ref{discuss_jeans}).

The ratio between gravitational and thermal energy is kept constant, and thus clouds with higher temperature have higher initial density. 
The canonical run T10M0+ is taken from the model C in Paper I with the ratio between free-fall time and sound-crossing time $t_{\rm ff} / t_{\rm sct} = 0.05$. 
The clouds are around virial equilibrium with the ratio between free-fall time and turbulence-crossing time $t_{\rm ff} / t_{\rm vct} = 1.1$. 
The numerical resolution $l_{\rm max}$, that specifies the smallest cells size, $1/2^{l_{\rm max}}$ of the total simulation box, 
is chosen to have reasonably comparable physical resolutions. 
Higher density threshold of $n_{\rm sink} = 3 \times 10^{11} ~\cc$ instead of $10^{10}$ is used for sink particle formation to avoid artificial collapse before the gas reaches the adiabatic part of the EOS. 
This is necessary for the high temperature runs, 
while we have demonstrated in Paper I that varying $n_{\rm sink}$ does not significantly affect the stellar mass spectrum. 
The parameters are listed in Table \ref{table_T_params}. 

\subsubsection{Polytropic index of the diffuse gas}
The diffuse ISM is often reasonably approximated by an isothermal gas. 
In this study, we introduced a small dependence on the density with the polytropic index $\gamma_0$ for the diffuse gas.  
The values of $T_0=10~{\rm K}$ and $\gamma=1.66$ are used, and the canonical run has $\gamma_0=1$ by construction. 
We performed two runs with $\gamma_0 = 1.2 ~{\rm and}~ 0.7$, respectively. 
The parameters are also listed in Table \ref{table_T_params}. 

The more diffuse gas is linked to the formation of cores of higher masses. 
With these setups, we aim to investigate whether different $\gamma_0$ values have an impact on the distribution of the diffuse gas, 
and consequently affect the high-mass end of the stellar mass spectrum. 
\\

\begin{figure}
\includegraphics[width=.5\textwidth]{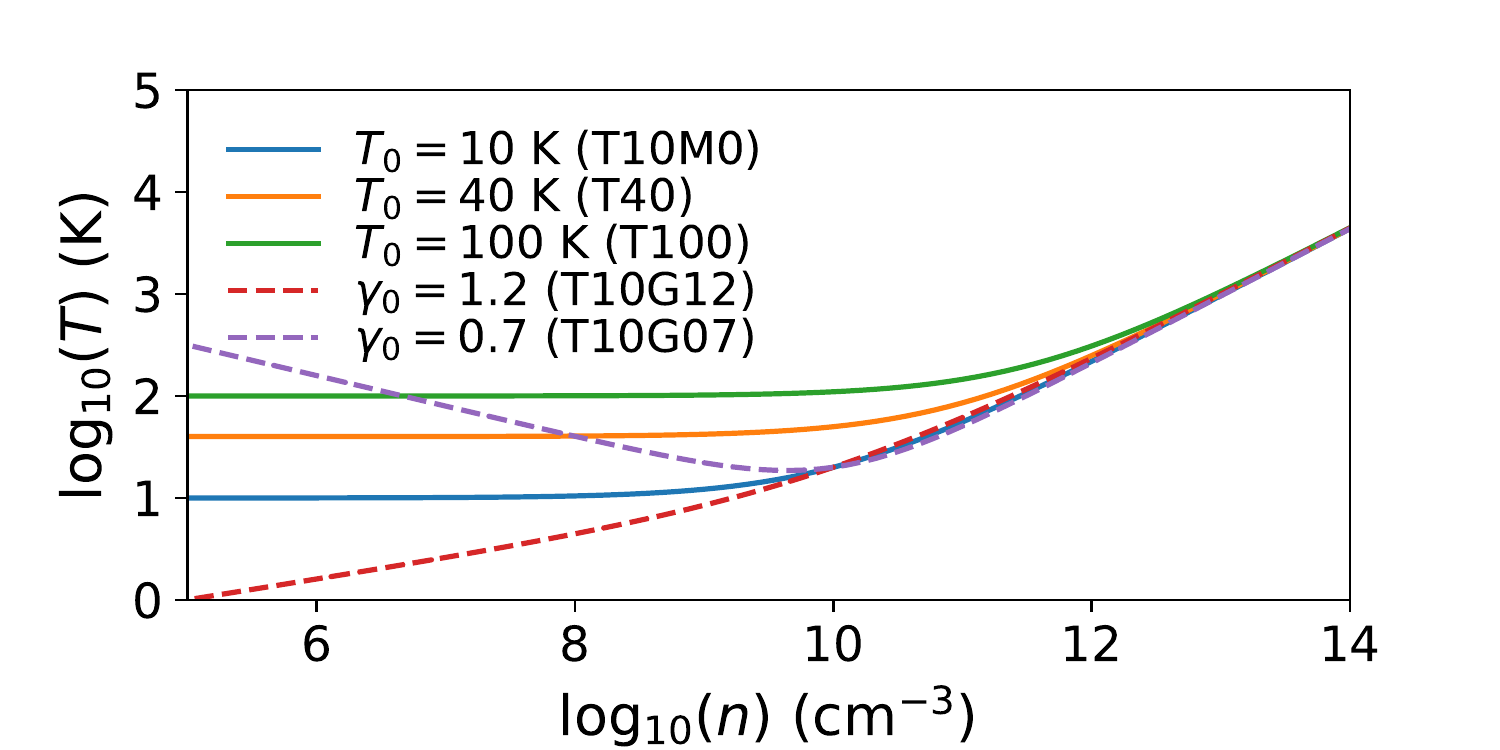}
\caption{The EOS plotted in temperature as function of gas number density, as listed in Table \ref{table_T_params}. Either the isothermal bulk temperature, $T_0$ (solid curves), or the polytropic index, $\gamma_0$ (dashed curves), is varied. The varied parameter and the corresponding run label are shown in the legend. With the purpose to understand the effects of temperature in low density gas on the fragmentation outcome, the high-density end of all EOS is chosen to coincide. }
\label{fig_EOS_all}
\end{figure}

The EOS applied in simulations of this section are presented in Fig. \ref{fig_EOS_all}. 
Either the isothermal bulk temperature or the polytropic index at low densities is varied, while the high-density end of all EOS is identical. 

\subsection{The magnetic field}
\begin{table*}[]
\caption{Parameters for magnetized simulations. The label of the run, the ratio between free-fall time and alfv\'en crossing time, the ratio between free-fall time and sound crossing time, the central density, the plasma $\beta$, the alfv\'enic Mach number, the maximal level of refinement, and the physical resolution are listed. }
\label{table_B_params}
\centering
\begin{tabular}{l c c c c c c c c c c}
\hline\hline
Label   & $t_{\rm ff} / t_{\rm act}$  & $t_{\rm ff} / t_{\rm sct}$ & $n_0$  & $\beta$ & ${\cal M}_{\rm A}$ & $l_{\rm max}$    & resolution (AU) \vspace{.5mm}\\
\hline
T10M0+   & $ 0 $ & $ 0.05 $ & $6.0 \times 10^7$ & inf& inf&15 & 2   \\%Lowmass_compact_relax_l8_2                                            
M01   & $ 0.1$ & $ 0.05 $  & & 0.5 & 11 &14 & 4   \\%Lowmass_compact_relax_l8_mag01_3                                           
M04   & $ 0.4 $ & $ 0.05 $ & & 0.03 & 2.75 &14 & 4   \\%Lowmass_compact_relax_l8_mag04_6                                          
M08   & $ 0.8$ & $ 0.05 $ && 0.008 & 1.375 &14 & 4   \\%Lowmass_compact_relax_l8_mag08_8
M12   & $ 1.2$ & $ 0.05 $ && 0.003 & 0.92 & 14 & 4   \\%Lowmass_compact_relax_l8_mag12_10       
\hline
T10M0A+  & $ 0$ & $ 0.15 $ &$8.2 \times 10^4$  & inf  &  inf &15 & 8   \\%Lowmass_isothermal_relax_l8_2_res                             
M04A  & $ 0.4$ & $ 0.15 $ && 0.03 & 2.75 & 14 & 17   \\%Lowmass_isothermal_relax_l8_mag04_6                                           
\hline
\end{tabular}
\end{table*}

The role of magnetic field during the fragmentation is also studied. 
The magnetic field provides extra support aside from the thermal pressure and turbulence, while also introducing some anisotropy. 
We varied the initial magnetic field strength to study the influences on the mass spectrum. 

With respect to the canonical run T10M0+, a magnetic field is added at several strength while keeping all the other parameters identical. The simulation is evolved following ideal magnetohydrodynamic (MHD) equations. We also perform a magnetized run  with respect to the case A (here labeled T10M0A+) from Paper I, which has a lower initial density and higher thermal to turbulent energy ratio. The parameters are listed in Table \ref{table_B_params}.

\subsection{Sink particles}
Sink particles are used to follow the high density self-gravitating regions below the resolution limit \citep{Bleuler14}. The sinks are formed at the highest level of refinement when the local density maximum exceeds the prescribed threshold, $n_{\rm sink}$, and the local gas is virially bound with converging flow. At the resolution of a few AU, it is guaranteed that a sink particle reasonably represents an individual star. Readers are invited to refer to Paper I for more detailed description of the algorithm and studies of numerical convergence. 

For simulations with higher $T_0$, a higher values of $n_{\rm sink}$ is used because $n_{\rm ad}$ is increased. 
This avoids spurious sink particles formation before the adiabatic temperature increase. 
Although the selection of $n_{\rm sink}$ is indeed slightly delicate, by varying the value of $n_{\rm sink}$, 
we have demonstrated in Paper II that this has no consequences on the characteristic mass of the mass spectrum.

\section{Simulation results}

\subsection{Sink particle mass spectrum}

\setlength{\unitlength}{1cm}
\begin{figure}[h!]
%\centering
\begin{picture} (0,18.7)
\put(0,12.6){\includegraphics[width=8.7cm]{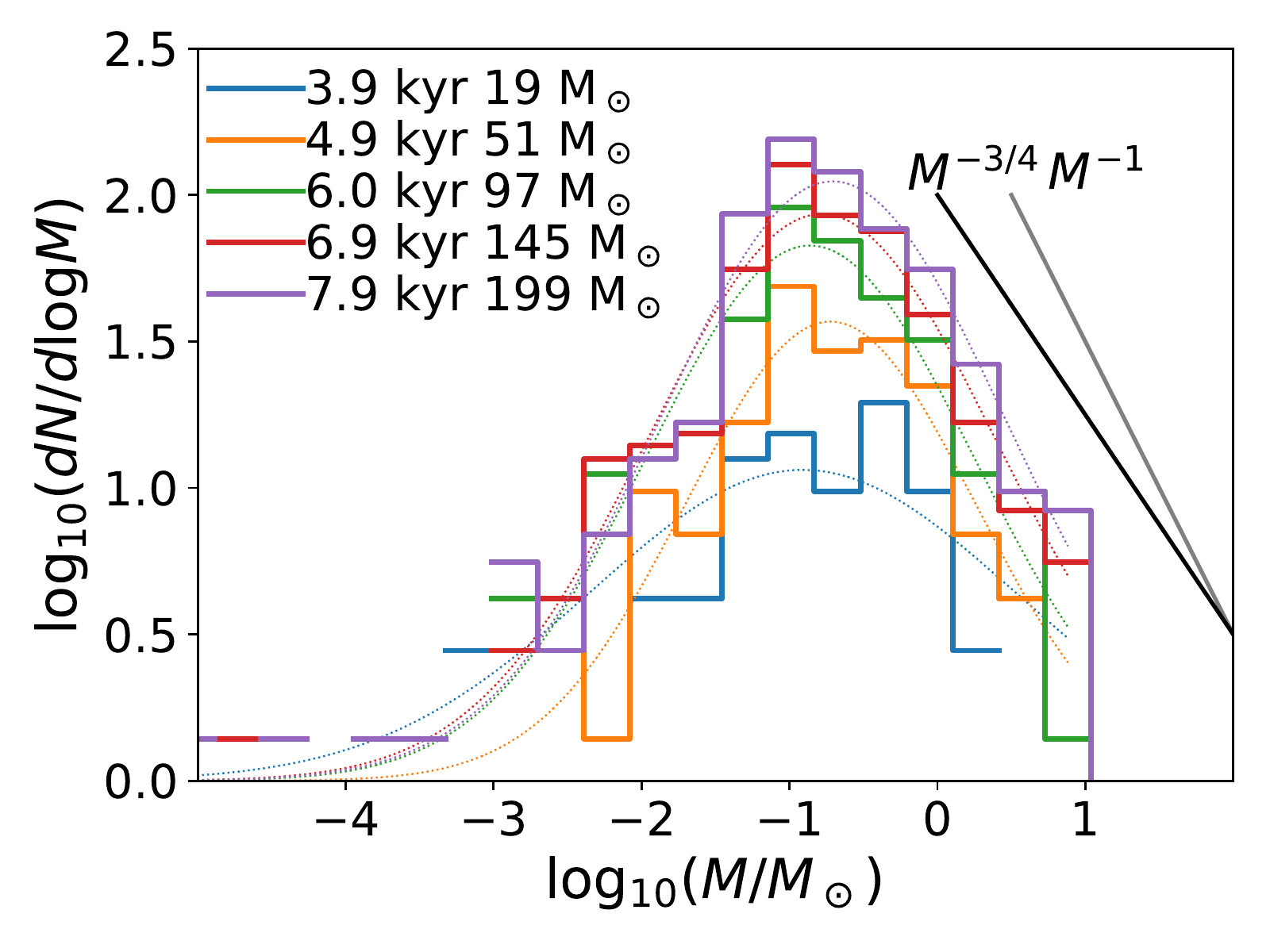}}  
\put(6.,18.4){T10B0+, 2 AU}
\put(0,6.3){\includegraphics[width=8.7cm]{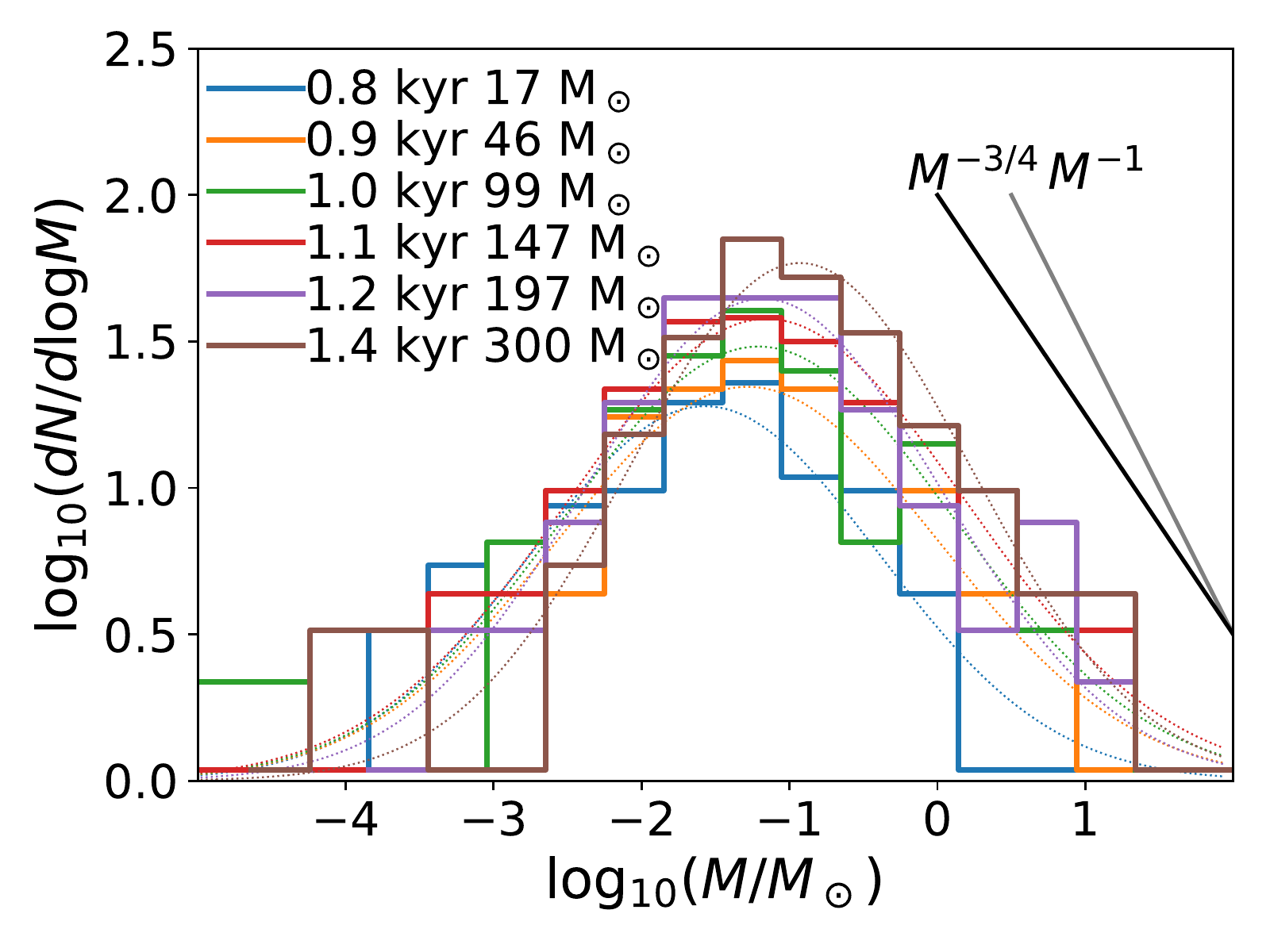}}  
\put(6.5,12.1){T40, 4 AU}
\put(0,0){\includegraphics[width=8.7cm]{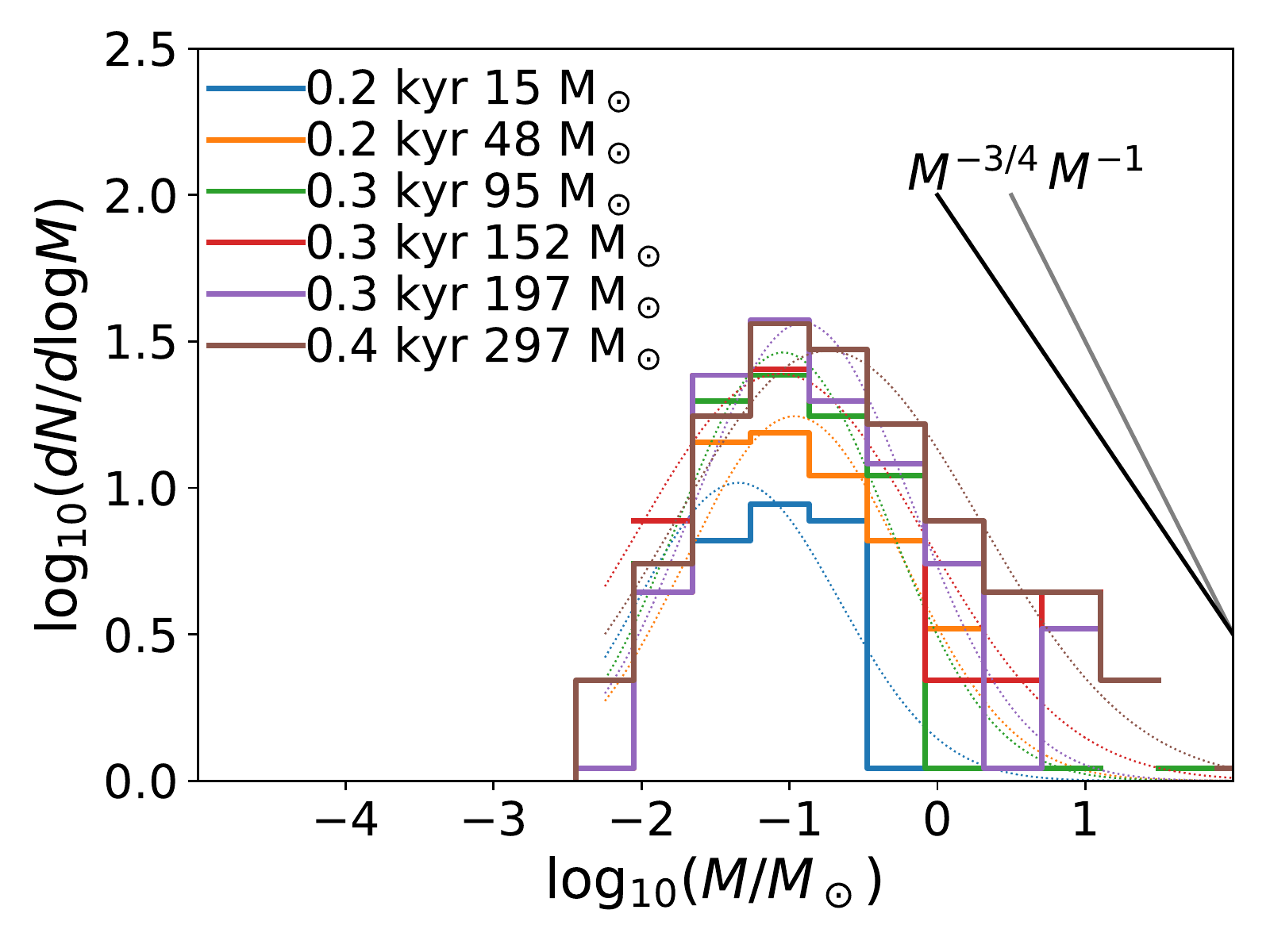}}  
\put(6.5,5.8){T100, 3 AU}
\end{picture}
\caption{Mass spectra with varying cloud bulk temperature. The colors blue, orange, green, red, purple, and brown represent the time at which about 20, 50, 100, 150, 200, and 300 $\Ms$ has been accreted onto sink particles. Log-normal fits are shown with thin lines for reference. The top panel is the canonical run T10M0+ (run C++ in Paper I), where the non-magnetized run has  an EOS that sets the low density gas to $T_0 =10$ K. 
Middle panel (T40) with $T_0=40$ K  shows similar peak position as the canonical run.
Lower panel (T100) with $T_0=100$ K also has same characteristic peak mass, while the formation of low-mass sinks is clearly more difficult. }
\label{fig_C_temp}
\end{figure}

\setlength{\unitlength}{1cm}
\begin{figure}[h!]
%\centering
\begin{picture} (0,12.4)
\put(0,6.3){\includegraphics[width=8.7cm]{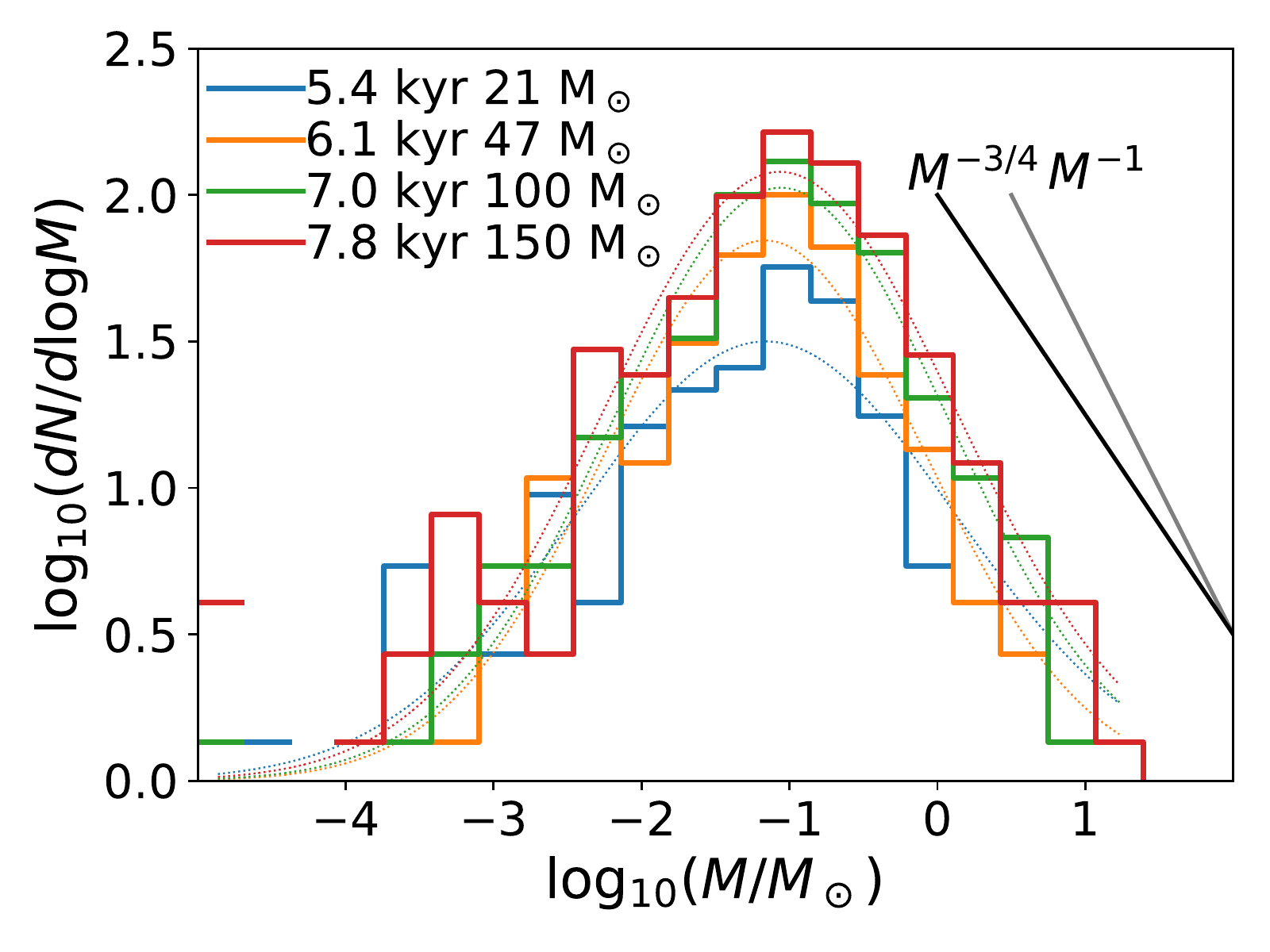}}  
\put(6.,12.1){T10G12, 4 AU}
\put(0,0){\includegraphics[width=8.7cm]{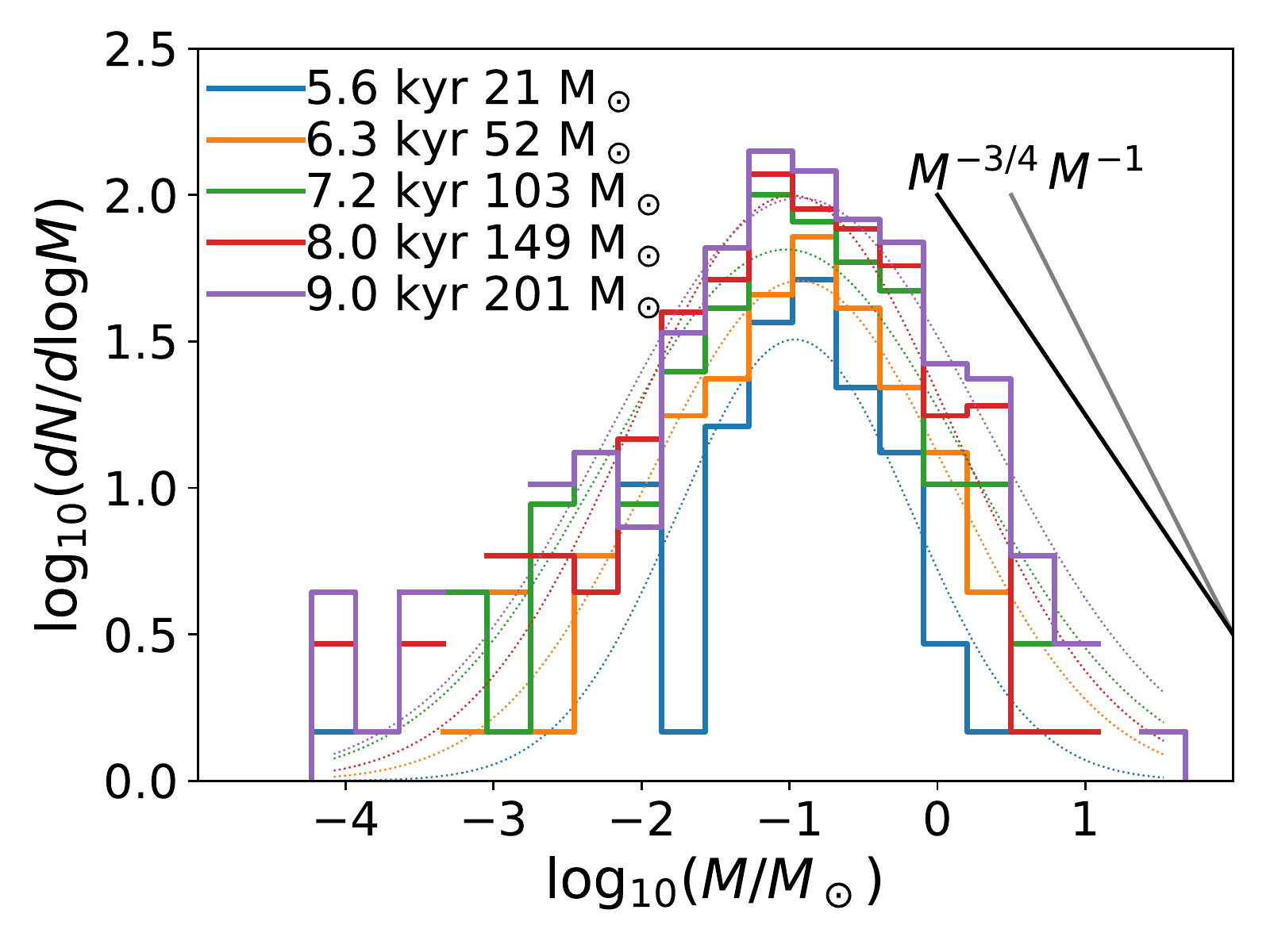}}  
\put(6,5.8){T10G07, 4 AU}
\end{picture}
\caption{Mass spectra with varying polytropic index $\gamma_0$ at low densities, same color coding as in Fig. \ref{fig_C_temp}. Log-normal fits are shown with thin lines for reference. Top panel has $\gamma_0=1.2$ and lower panel has $\gamma_0=0.7$. The mass spectra are very similar to those of the canonical run, and mostly importantly the peak position is not altered. The high-mass end slope of run T10G12 seems to be slightly steeper, and that of T10G07 shallower in the mass range from $0.1$ to $1~\Ms$.  }
\label{fig_C_gamma}
\end{figure}

The mass spectra of sink particles are shown in Fig.~\ref{fig_C_temp} for simulations with varied bulk temperature, $T_0$. 
The three panels from top to bottom show runs with $T_0 = 10,~40,~{\rm and}~ 100~{\rm K}$.  
The mass spectra are plotted at times when $20, ~50, ~100, ~150, ~200, ~{\rm and}~ 300~\Ms$ are accreted, where applicable since some runs are more evolved than the others. 
The number of stars decreases  with increasing $T_0$, 
and the run T100 formed a few more massive stars. 
This is probably because the global density in run T100 is higher. 
In consequence, the fluctuations reach the adiabatic branch of the EOS more easily, 
and small fragments form less aboundantly.
On the other hand, the overall shape of the mass spectrum does not seem to vary much from one simulation to another. 
The peak at $\sim 0.1~\Ms$ is almost unchanged for the three runs of Fig.~\ref{fig_C_temp}.

Figure \ref{fig_C_gamma} shows the two runs with $\gamma_0 = 1.2~{\rm and}~0.7$. 
There is no remarkable difference with respect to the canonical run. 
One might notice a slightly steeper high-mass end slope for the run T10G12, 
and a slightly shallower slope for T10G07 in the range $0.1-1~\Ms$. 
We discuss this as a possible effect of different values of $\gamma_0$ in Sect. \ref{st_dicuss_gam0}.

Figure \ref{fig_C_mag} shows mass spectra from simulations with varied initial magnetic field strength with respect to the canonical run T10B0+. 
From run M01, M04, to run M08, there seems to be a flattening around the mass spectra peak and a slight broadening of the whole spectrum. 
Nonetheless, despite this weak trend, the spectra are very similar.  
Run M04 is not shown for conciseness since the spectra are also similar. 
The most magnetized run, M12, shows different behavior with more pronounced peak and limited number of low-mass stars. This may originate from the highly anisotropic geometry that resulted from the strong magnetic field.

Figure \ref{fig_A_mag} shows the run A from Paper I and a magnetized run at same density. 
As discussed in Paper I, the mass spectrum becomes almost flat when the initial thermal energy is relatively important compared to the turbulent energy (i.e. lower density). 
The introduction of magnetic field makes the mass spectrum more noisy possibly by making the collapsing flow anisotropic, 
and slightly more massive stars are formed while the number of low mass objects is reduced. 
Overall, we do not see obvious qualitative differences in the two runs. 

\setlength{\unitlength}{1cm}
\begin{figure}[h!]
%\centering
\begin{picture} (0,18.7)
\put(0,12.6){\includegraphics[width=8.7cm]{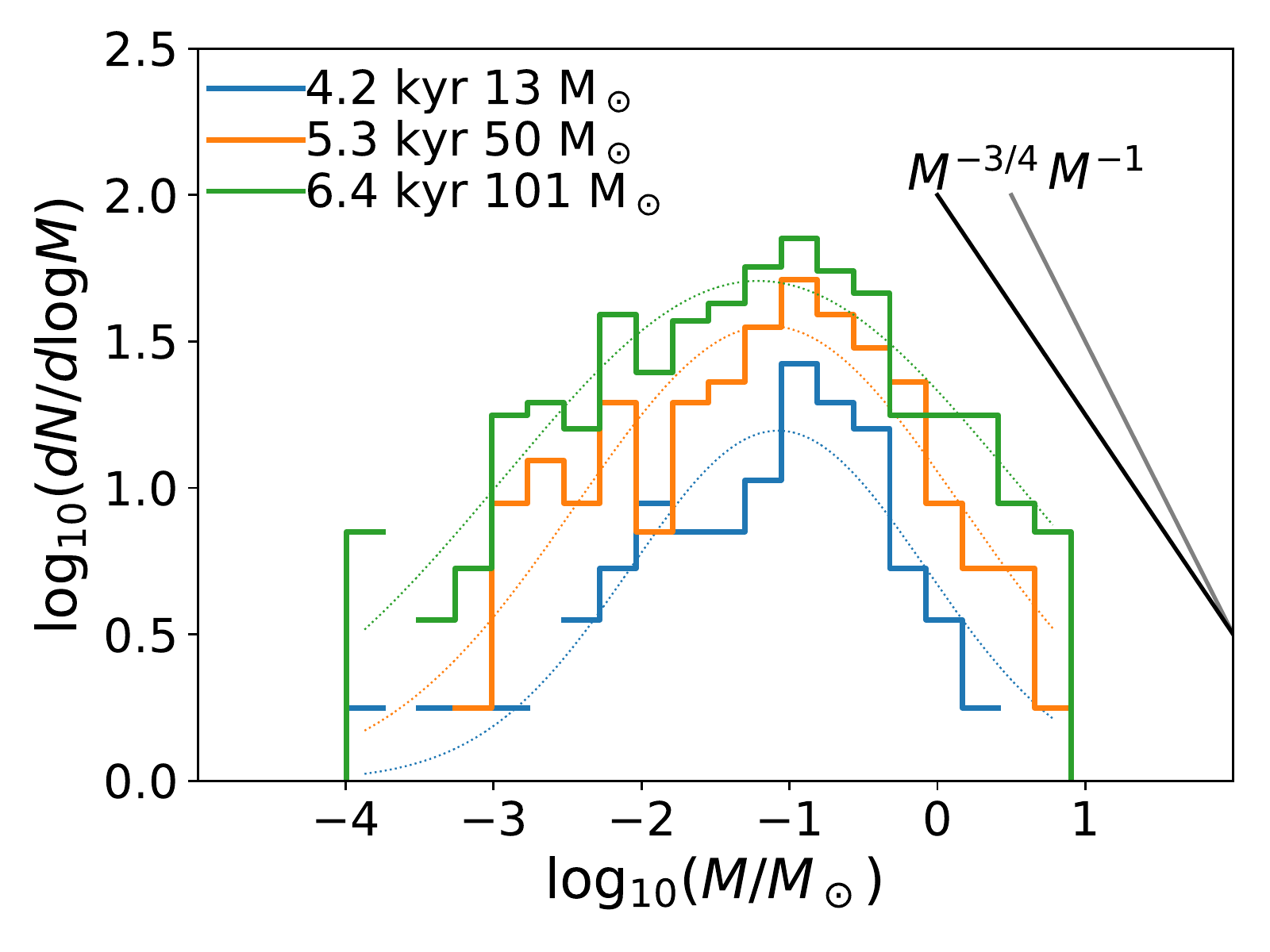}}  
\put(6.5,18.4){M01, 4 AU}
\put(0,6.3){\includegraphics[width=8.7cm]{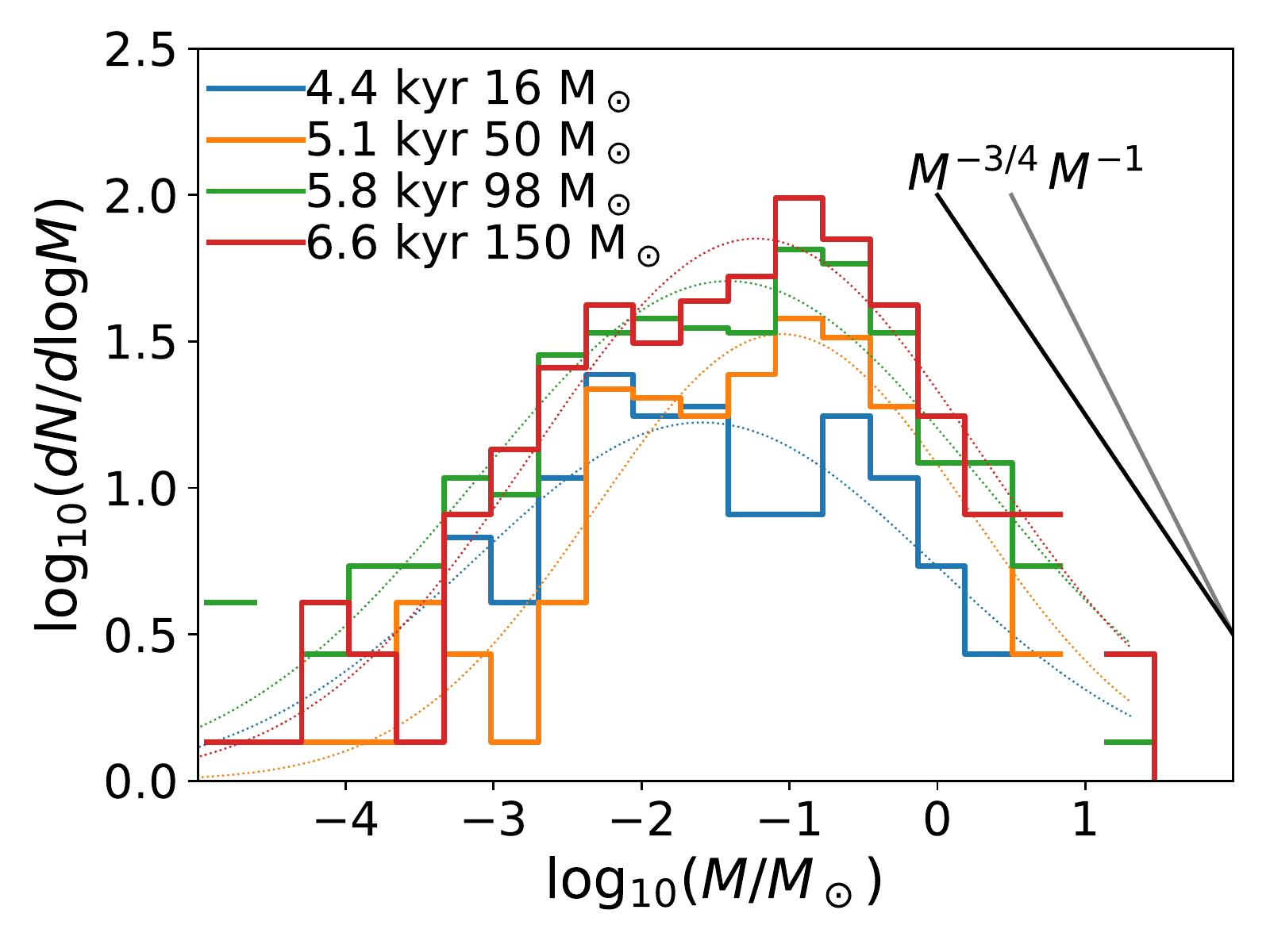}}  
\put(6.5,12.1){M08, 4 AU}
\put(0,0){\includegraphics[width=8.7cm]{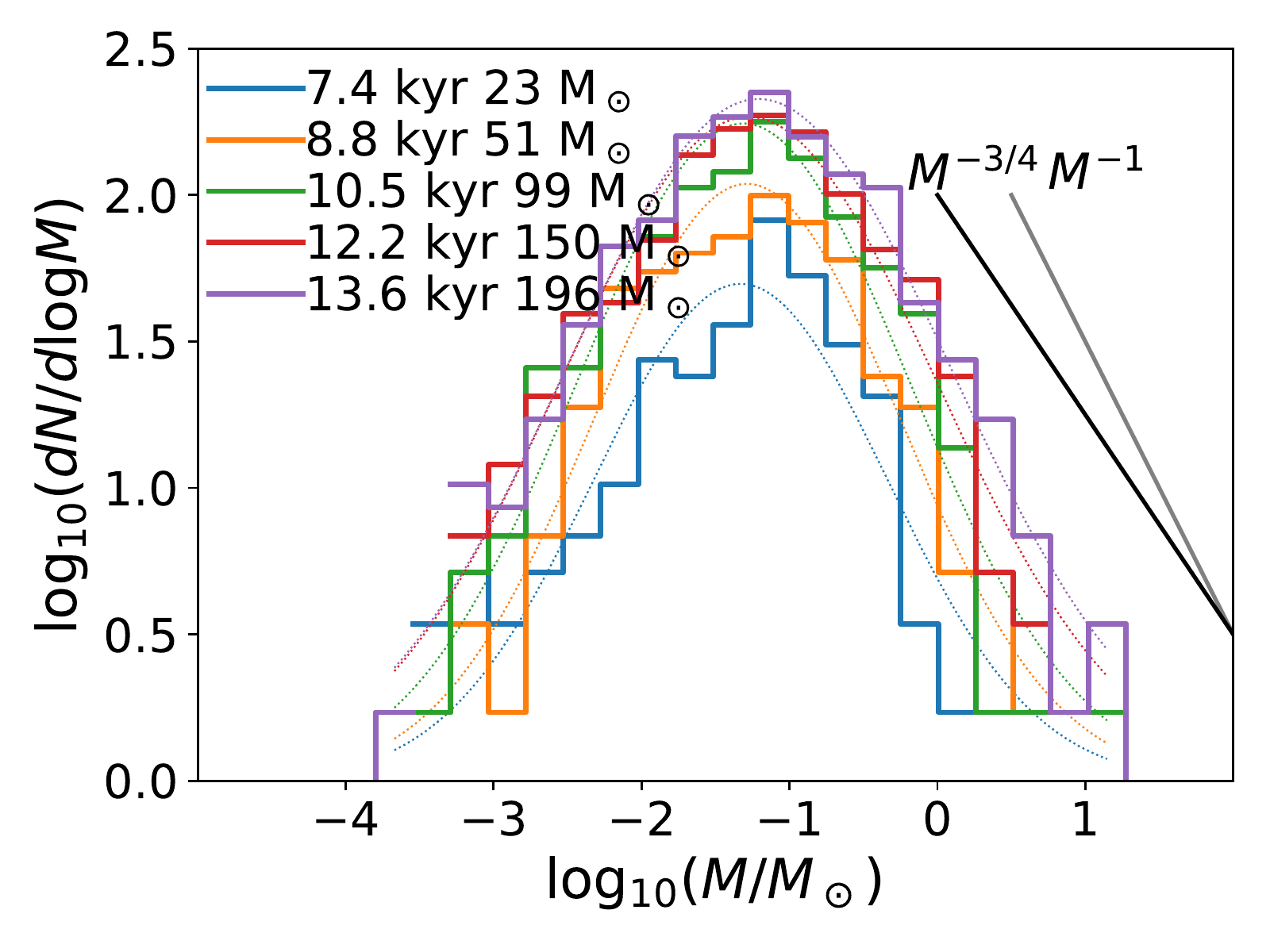}}  
\put(6.5,5.8){M12, 4 AU}
\end{picture}
\caption{Mass spectra with varying cloud magnetization, same color coding as in Fig. \ref{fig_C_temp}. Log-normal fits are shown with thin lines for reference. All runs have identical initial conditions as the canonical run except for the magnetic field strength that increases from top to bottom (M01, M08, and M12). The run M04 is not shown for conciseness since the spectra are not too different from either M01 or M08. 
The mass spectra are slightly broadened with respect to the non-magnetized runs. Nonetheless, the magnetic field provides extra support without affecting the characteristic mass of cloud fragmentation in general. Only the most magnetized run (M12) with extreme field strength exhibits slightly different behavior probably to the anisotropic cloud collapse.}
\label{fig_C_mag}
\end{figure}

\setlength{\unitlength}{1cm}
\begin{figure}[h!]
\begin{picture} (0,12.4)
\put(0,6.3){\includegraphics[width=8.7cm]{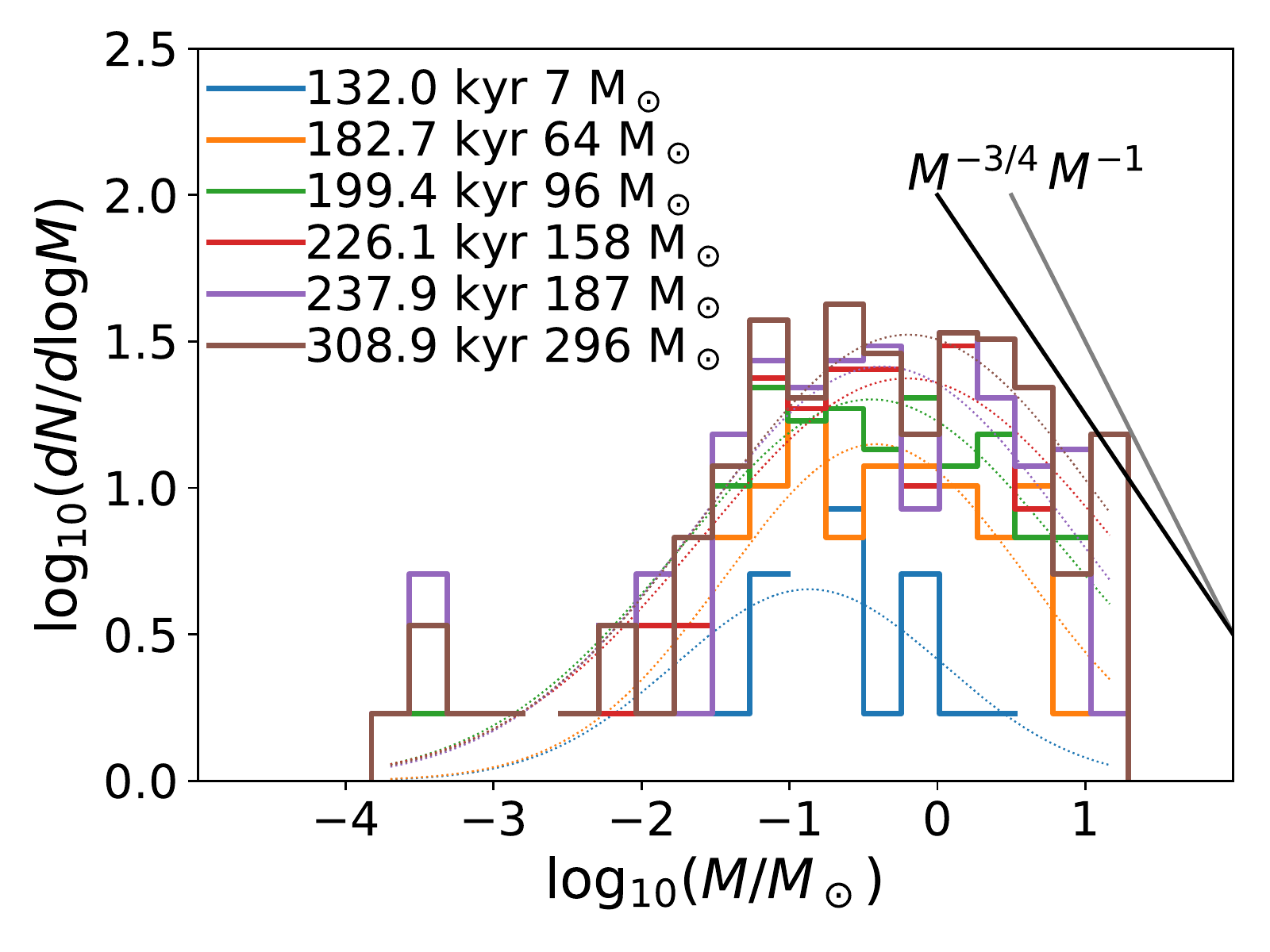}}  
\put(5.8,12.1){T10M0A+, 4 AU}
\put(0,0){\includegraphics[width=8.7cm]{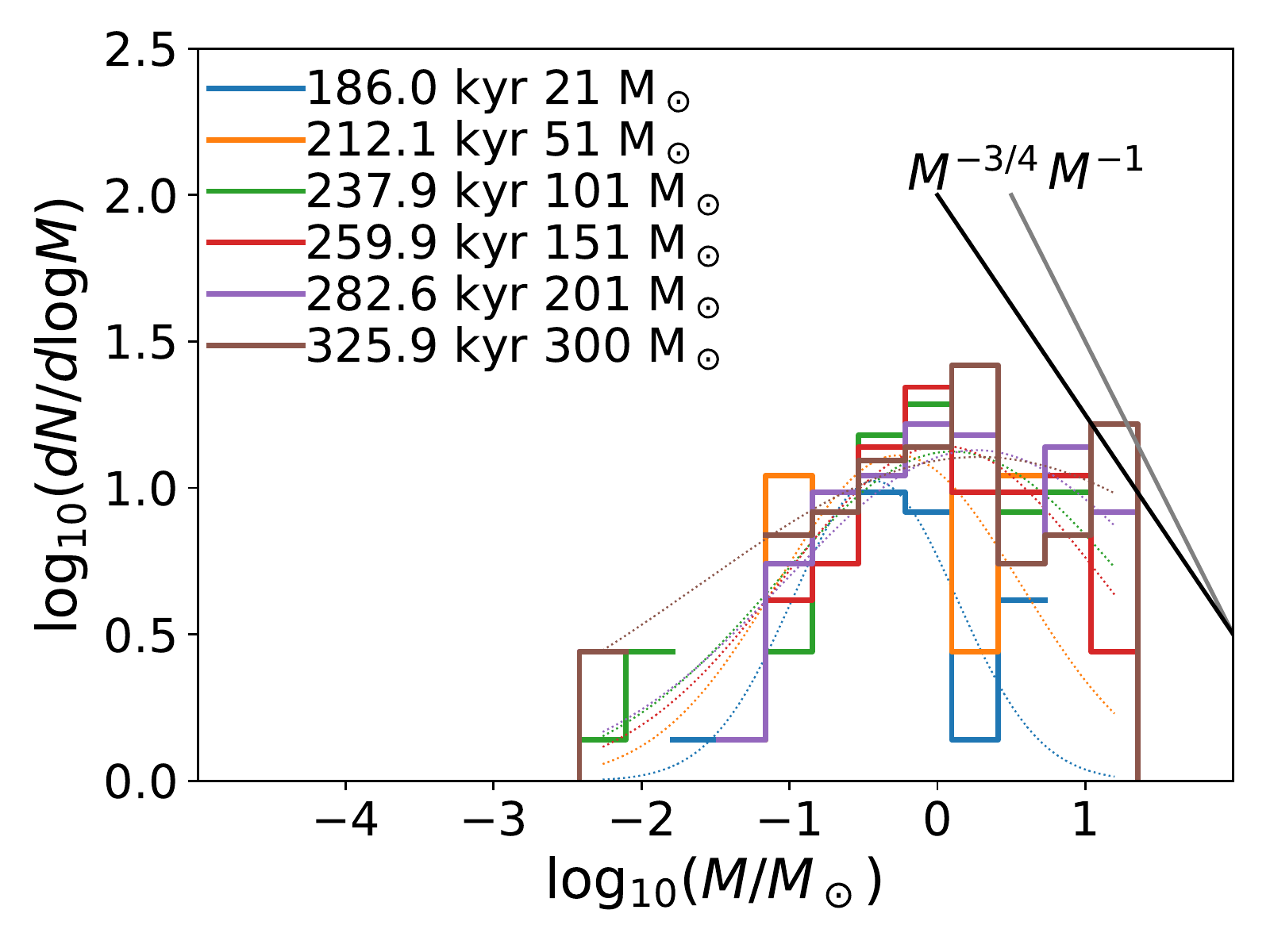}}  
\put(6.3,5.8){M04A, 4 AU}
\end{picture}
\caption{The top panel (T10M0A+) is the case A in Paper I, which has lower initial density compared to the canonical run. Same color coding as in Fig. \ref{fig_C_temp} is used. Log-normal fits are shown with thin lines for reference.  As explained in Paper I, the thermal support governs the mass fragmentations and the mass spectrum is almost flat. Introducing the magnetic field in the lower panel (M04A) does not change significantly the overall behavior by adding extra support. }
\label{fig_A_mag}
\end{figure}

\subsection{Density PDF}\label{st_PDF}
\begin{figure}
\includegraphics[trim=72 0 113 0,clip,width=.24\textwidth]{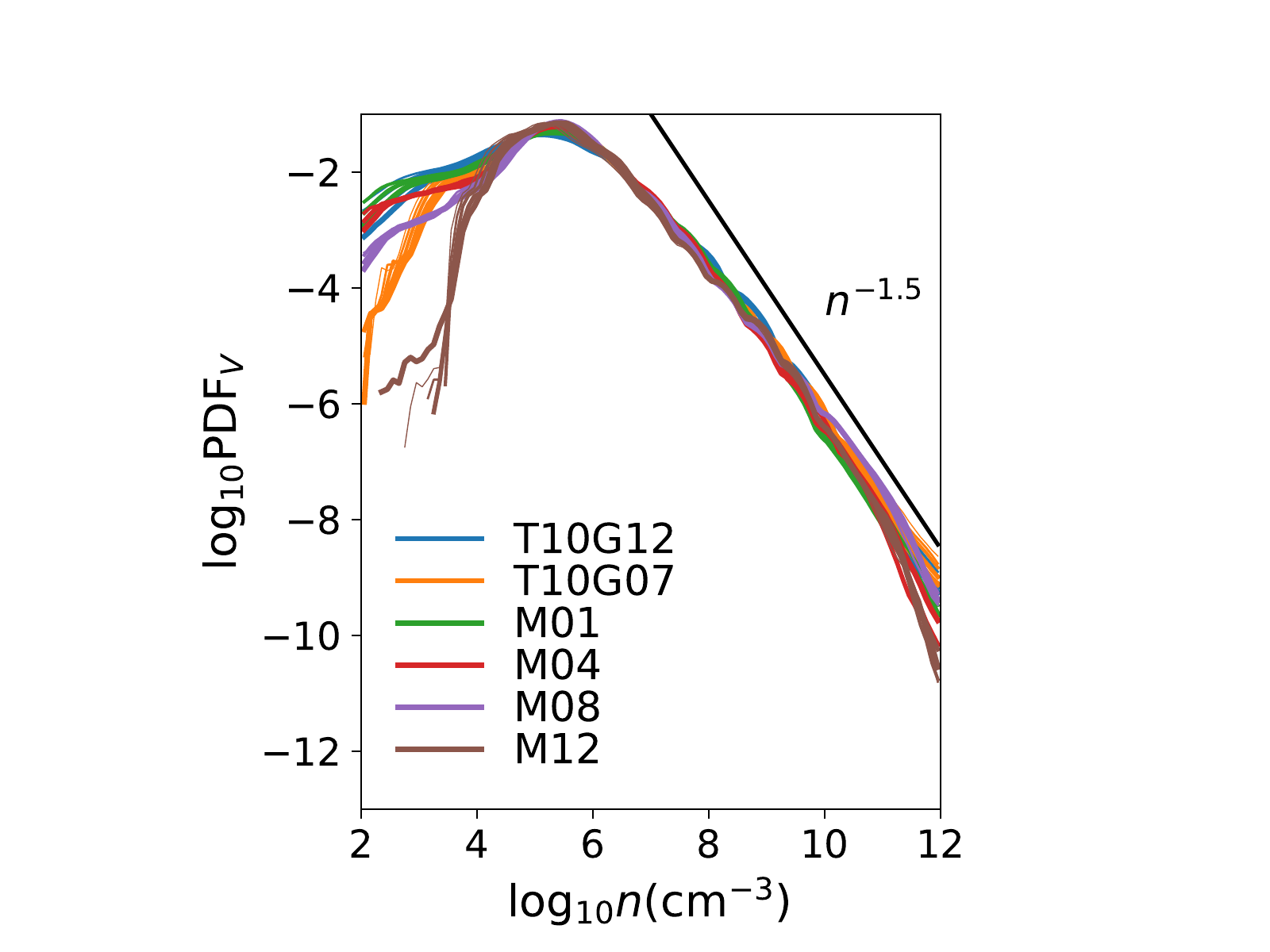}
\includegraphics[trim=72 0 113 0,clip,width=.24\textwidth]{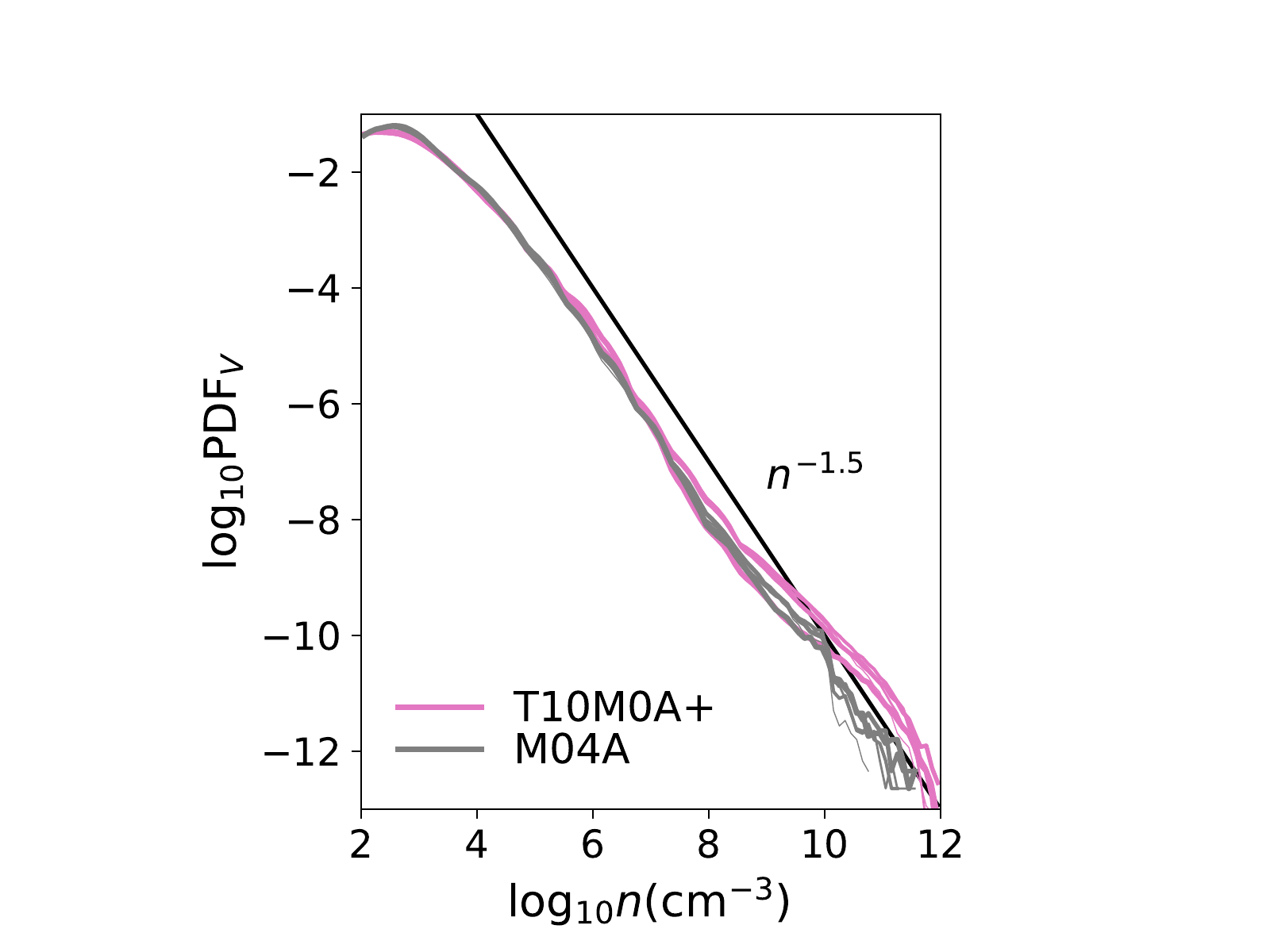}
\caption{{\it Left:} Density PDFs for runs with varied $\gamma_0$ and magnetization. The same time steps as used for the mass spectra are shown with decreasing line width. The density PDF is almost invariant in time. The high-density end merely changes from on run to another. A powerlaw of $n^{-1.5}$ is shown for comparison. The low-density end is narrower in run T10G07 because of the increasing thermal pressure with decreasing density. More magnetized runs have less low density gas as well, due to similar effects. {\it Right:} Density PDF of model A (lower initial density) without and with magnetic field. The magnetic filed does not affect the PDF in the density range where stars are actively forming.  }
\label{fig_PDF}
\end{figure}

The mass spectrum is closely linked to the density probability function (PDF) since it tells how much gas is subject to local collapse at each critical mass. 
We show in Fig. \ref{fig_PDF} the volume-weighted density PDF of the simulations and discuss the effect of varying the parameters. 
The left panel shows runs which has same initial density as the canonical run. 
The gas cools at low density for $\gamma_0=1.2$ (T10G12) and this has no visible effects on the density PDF. 
On the other hand, the thermal pressure increases with decreasing density for $\gamma_0=0.7$ (T10G07). 
This lowers the Mach number at fixed turbulence strength and the density PDF is thus narrower at the low-density end. 
As for the magnetized runs, the density PDF at low densities narrows with increasing magnetic field strength. 
This might be due to the magnetic tension that prevents the gas from expanding. 
At the same time the increased magnetic pressure also narrows the density PDF.  
The effect is most obvious with the run M12. 
Most importantly however, the high-density ends of these PDF do not differ much  from one another, 
and this is the most relevant part for the unvarying mass spectra. 
A powerlaw is plotted for comparison and this high-density end slope is invariantly compatible with $n^{-1.5}$. 
In the presence of self-gravity-driven turbulence, a collapsing cloud develops naturally a powerlaw density profile \citep{Murray15,Murray17}, 
and thus in turn a powerlaw density PDF.

\section{Discussion: physical effects on the mass spectrum}

\subsection{Effect of varying the temperature}
Two major effect from the thermal pressure are expected. 
First, the isothermal temperature provides a scale-independent support against self-gravity. 
Secondly, the thermal pressure also affects the gas density PDF, which in turn affects the number of dense fluctuations that are susceptible to collapse.

\subsubsection{Bulk temperature of the diffuse gas}
\label{discuss_jeans}
Our simulations are inspired by this argument and aims to clarify the effect of the temperature. 
The Jeans mass at $n_{\rm ad}$ in the simulations is 
\begin{align}
M_{\rm J, crit} &=  5.1 \times 10^{-4} \Ms \left({n_{\rm ad}\over 10^{10}~ \cc}\right)^{-1/2} \left({T_0 \over 10 ~{\rm K}}\right)^{3/2} \nonumber \\
&= 5.1 \times 10^{-4} \Ms  \left({T_0 \over 10 ~{\rm K}}\right)^{3/2 - 1/2/(\gamma-1)}  \nonumber \\
& \propto T_0^{3/4},
\end{align}
where the second equality results from the $n_{\rm ad}$ dependence that we chose in order to coincide the high-density end of the EOS. 
If this critical Jeans mass is the quantity determining the IMF peak, 
we should then see the peak of the mass spectra from our simulations shifting almost linearly with $T_0$, 
that is, $0.00051, 0.0014, {\rm and}~ 0.0028 ~\Ms$ for $T_0 = 10, ~40, {\rm and}~ 100 ~{\rm K}$. 
Contrarily, the peak mass in our simulations is independent of $T_0$ and situates at $\sim 0.1 ~\Ms$ (see Fig. \ref{fig_C_temp}), 
which is two orders of magnitude larger than the critical Jeans masses. 
As suggested in Paper II, the characteristic mass of the IMF is indeed determined by the high-density end of the EOS, 
which is identical in this set of simulations. 
The lack of variation of the peak mass confirms this result. 

With the gas heating up adiabatically, the first Larson core, a pressure-supported hydrostatic core, forms 
and it has a  mass  $M_{\rm L} \sim 0.02~\Ms$. 
In paper II, we have derived the mass of the first Larson core by simply integrating the hydrostatic equilibrium equations while imposing the density and flat profile at the center as boundary conditions. When using an adiabatic EOS and sufficiently high central density, the first Larson core is always well-defined with a steep density drop at a few tens of AU. This density profile defines a mass, that is not very sensitive to the central density. Such calculations give an idea of the typical mass of the first Larson core before the formation of the central singularity, the protostar. 

We caution that the exact value of this mass depends on the detailed radiation-thermodynamics \citep{Masunaga98,Vaytet17} and our treatment by a barotropic EOS is indeed very simplistic. 
The final mass of the star is determined by the competition between the fragmentation inside the surrounding envelope that tends to form new stars and block accretion and the tidal field of the star and envelope that tends to shear out density fluctuations and favor accretion. 
The result is a characteristic mass at $\sim 10~M_{\rm L}$ \citep[][Hennebelle et al 2018 submitted]{Lee18b}.  
 
\subsubsection{Polytropic index of the diffuse gas}\label{st_dicuss_gam0}
Following the previous discussions, it is already expected that the polytropic index of the diffuse gas, 
$\gamma_0$, in Eq. (\ref{eq_EOS2}) has no obvious effects on the peak value of the mass spectrum. 

The density PDF, as already shown in Sect. \ref{st_PDF}, does not depend on $\gamma_0$ at high densities. 
The powerlaw index with invariant value of -1.5 is probably an outcome of the turbulent global collapse and independent of $\gamma_0$. 
We discuss the effect of $\gamma_0$ in Appendix \ref{append_gam0} if it were to be taken into account in the density PDF.
Asides from this, here we discuss the impact of $\gamma_0$  in the thermally supported regime.

First as a reminder, the gravo-turbulent fragmentation model of CMF \citep{HC09} was applied in Paper I to account for the powerlaw slope of the mass spectra while the density PDF is replaced by a powerlaw. 
The virial equation of mass supported by dominating turbulent energy gives (Eq. (11) of Paper I)
\begin{align}\label{eq_M_R}
M \propto R^{1+2\eta} ~~~\text{and}~~~ \rho \propto M^{(2\eta-2)/(1+2\eta)}, 
\end{align}
where $R$ is the size of the self-gravitating structure and $\eta=0.5$ is the index of the turbulence power spectrum. 
Taking Eq. (2) from Paper I and density PDF $\propto \rho^{-p}$, we get the mass spectrum of self-gravitation structures: 
\begin{align}\label{eq_spec_mass_T}
{dN  \over d\log M}  &= M{\cal N} (M)  
\propto {\rho \over M} {\cal P}(\rho) \propto M^{[-3+p(2-2\eta)]/(1+2\eta)}.
\end{align}
The mass spectrum has powerlaw index $-3/4$ if $p=1.5$ is applied. 
We call this the TG regime (T and  G for turbulent supported and global density PDF), which is what we see in most cases.

When the thermal support is dominant, then the mass-size relation from virial equilibrium becomes  
\begin{align}
M \propto R^{(4-3\gamma_0) / (2-\gamma_0)} \propto \rho^{(3\gamma_0-4)/2},
\end{align}
which leads to
\begin{align}\label{eq_spec_mass_P}
{dN  \over d\log M}  &
\propto {\rho \over M} {\cal P}(\rho) 
\propto M^{[2(p-1)+3\gamma_0-4]/ (4-3\gamma_0)}
\propto M^{-3(1-\gamma_0)/ (4-3\gamma_0)}, 
\end{align}
where the last result is obtained by applying $p=1.5$
We call this the PG regime (P for pressure).
The powerlaw index of the mass spectrum is 0 for $\gamma_0=1$, as already know from paper I, 
and it becomes -0.47 with $\gamma_0=0.7$. 
This is probably what we see in run T10G07 in the mass range between 0.1 and 1 $\Ms$ with slightly shallower slope since the thermal support is stronger than that in the canonical run at low densities and the thermally supported regime emerges. 
The higher masses might reach the regime where the density PDF becomes lognormal and thus have a steeper slope.

Though from the density PDF and the model consistency it is mostly reasonable to use $p=1.5$ for the density PDF, 
we still do similar calculations for a local thermal density PDF in Appendix \ref{append_gam0} for comparison (regimes TL and PL). 
The resulting powerlaw index of the mass spectrum is listed in Table \ref{table_powerlaw} for the $\gamma_0$ values in our runs. 
As a reminder, the discussed cases correspond to the density regime II in Paper I, where the gravity makes the density PDF a powerlaw. 
The labels P and T correspond to support regimes a and b, correspondingly. 
The case of turbulent support combined with the global density PDF (TG) is simply the same as the TL case with $\gamma_0=1$ since $\gamma_0$ does not enter anywhere in the calculations. 
Meanwhile, the origin of this powerlaw index $p=1.5$ still remains to be further investigated. 

\begin{table}[]
\caption{Powerlaw index of the mass spectrum in different model regimes. The labels indicate the supporting agent (Turbulence/Pressure) and the PDF (Global/Local).}
\label{table_powerlaw}
\centering
\begin{tabular}{l c c c}
\hline\hline
$\gamma_0$  & 0.7  & 1 & 1.2 \\
\hline
TG & - & -0.75 & -\\
PG &  -0.47 & 0 & 1.5 \\    
TL($\eta=0.5$) & -0.52 & -0.75 & -0.90 \\
PL & 0 & 0& 0\\                     
\hline
\end{tabular}
\end{table}

\subsection{The impact of the magnetic field}
Recalling from Paper I that the mass spectrum $\mathcal{N}(M)  = dN/d\log M \propto M^0$ when the thermal energy dominates over turbulence 
and $\propto M^{-3/4}$ otherwise, the magnetic field field seems to behave slightly similar to the thermal pressure. 
The mass spectra of the magnetized runs are wider with less remarkable peak and show a slightly flattened shape (between 0.01 and 1 $~\Ms$) than the non-magnetized one. 
Nonetheless, the variation among these runs are not very strong, except for run M12 with very strong initial field. 
The whole cloud collapses along the field lines and has a more flattened geometry. 

\begin{figure}
\includegraphics[width=.5\textwidth]{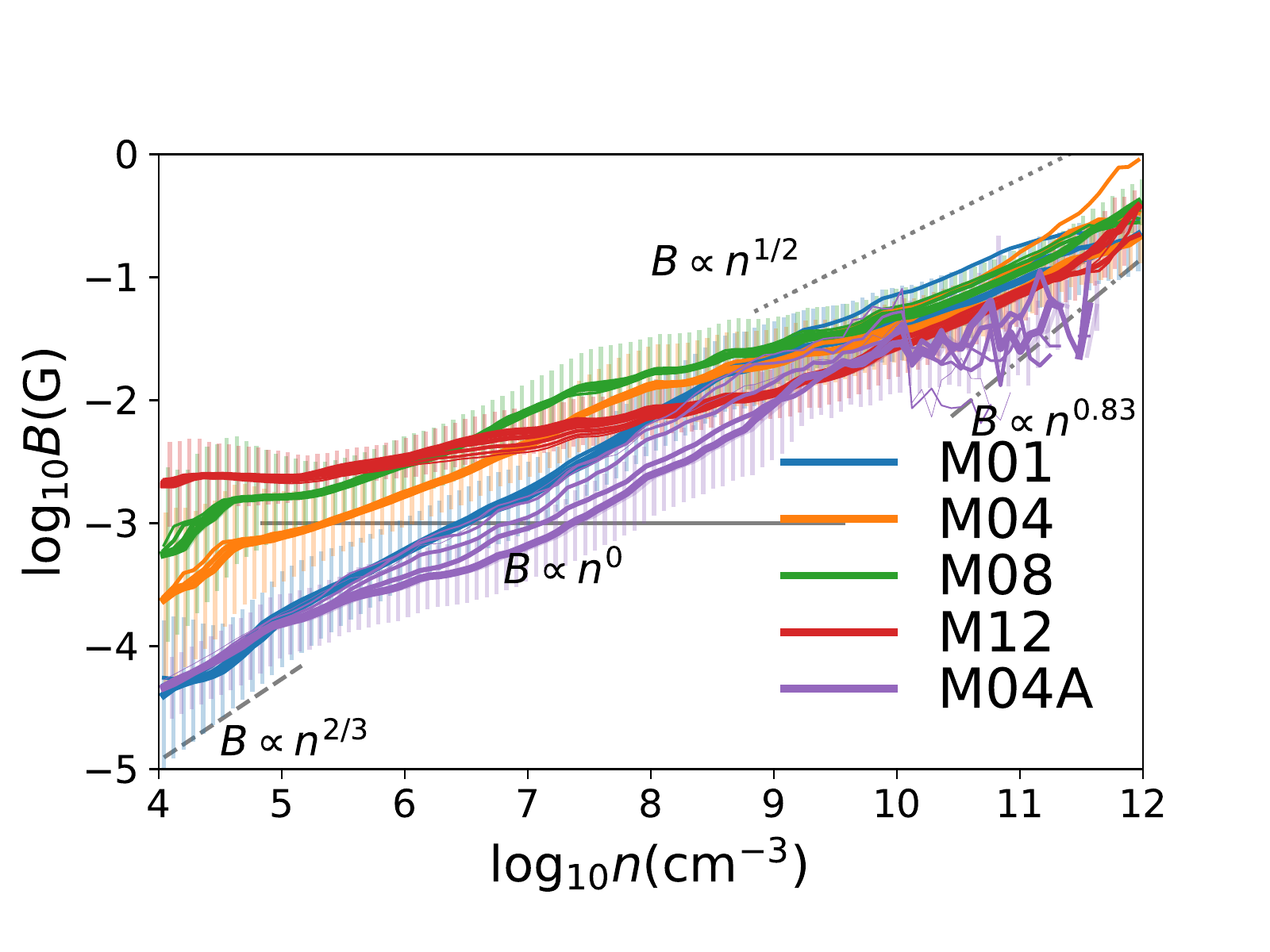}
\caption{The averaged magnetic field strength against density from runs M01, M04, M08,  M12, and M04A in blue, orange, green, red, and purple. The same time step as in Fig. \ref{fig_C_mag} and plotted with line width that decreases with increasing in time. Standard deviations are shown with shades. The relations $B\propto n^{2/3}$ (dashed), $B\propto n^{0}$ (plain), $B\propto n^{1/2}$ (dotted), and $B\propto n^{0.83}$ (dot-dashed) are plotted for reference. }
\label{fig_B_rho_all}
\end{figure}

The statistics of the magnetic field strength are plotted against density for the four runs in Fig. \ref{fig_B_rho_all}. 
The runs M01, M04, M08, M12, and M04A are plotted in blue, orange, green, red, and purple respectively. 
The lines with decreasing width correspond to the time steps where the sink mass spectra are shown in Fig. \ref{fig_C_mag}, 
and the shadows show the logarithmic standard deviation. 
The averaged magnetic field-density relation is almost invariant in time. 
With some dispersions, there is a clear trend of the field strength increasing with increasing density. 

Three regimes are observed. 
We discuss  them using the relative importance of thermal (T), gravitational (G), and magnetic (B) energies. 
This is a simplified view since we only compare at global scales and there lacks real scale arguments. 
Various density dependences are plotted for reference. 

\begin{itemize}
\item G>B>T:
At low densities ($< 10^5 ~\cc$) with weak magnetic field, 
the gravitational contraction proceeds isotropically without seeing the field. 
Simple magnetic flux freezing arguments lead to the relation $B \propto n^{2/3}$ \citep{Mestel65}.

\item B$\sim$G>T: 
At intermediate densities with strong field, the $B-n$ relation flattens. 
This corresponds to the regime where the field is strong enough to dominate the mass flow while the density is still too low to provide thermal support. 
The mass piles up along the filed lines, and the density increases without increasing the magnetic flux.  
This regime only exists in runs with stronger initial field (most clearly in M12).

\item B$\sim$G$\sim$T:
The high-density ends of the four runs almost coincide, 
and this possibly explains the very weak variation of the mass spectra. 
The thermally supported contraction along field lines at constant $\beta = P_{\rm th} / P_{\rm B}$ leads to $B \propto n^{\gamma/2}$ \citep{Mouschovias91}.
The exponent becomes 0.5 and 0.83 with $\gamma=1 ~\text{and}~1.66$ for density below and above $10^{10}~\cc$. 

\end{itemize}

The behavior of the magnetic field is the outcome of the interplay between gravity, thermal pressure, and the filed itself. 
The second regime is more clearly seen in run M12 with a strong initial field, 
and this implies an important anisotropy introduced by the presence of the magnetic field. 
The runs with weaker magnetic fields probably go directly from the first regime to the third one. 

Interestingly, when the thermal pressure becomes strong enough to support against the magnetically-guided collapse at high densities in run M12, 
the field joins the $B\propto n^{\gamma/2}$ relation, 
but with a lower absolute value with respect to the other runs. 
The might be the reason for the different shape of mass spectrum of run M12, which is narrower with a rounder peak.

\begin{figure}
\includegraphics[width=.5\textwidth]{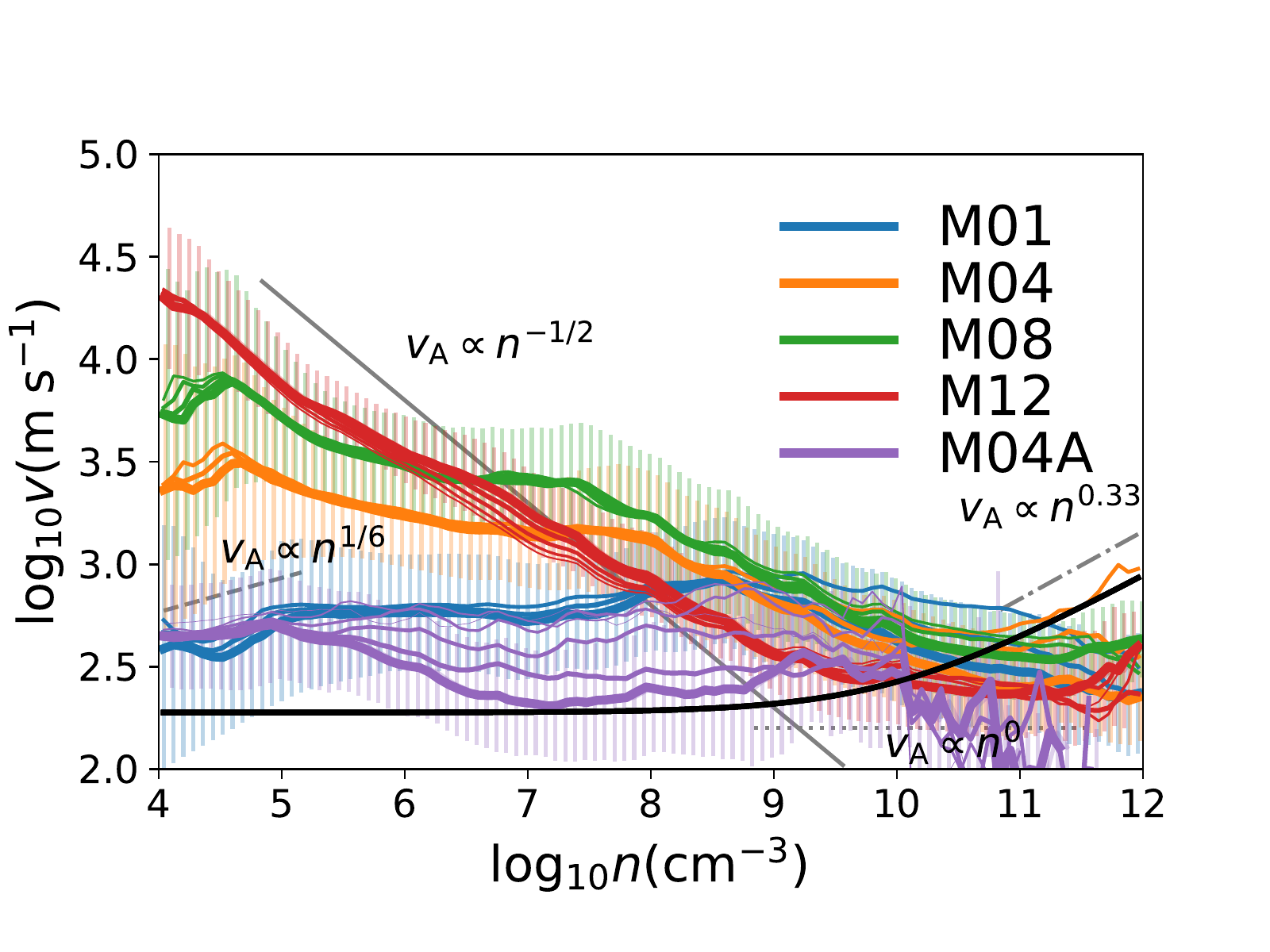}
\caption{The averaged Alfv\'en velocity against density from runs M01, M04, M08, M12, and M04A, in blue, orange, green, red, and purple. The same time step as in Fig. \ref{fig_C_mag} and plotted with line width that decreases with increasing in time. Log standard deviations are shown with shades. The thermal sound speed is plotted in black for comparison. The relations $v_{\rm A}\propto n^{1/6}$ (dashed), $v_{\rm A} \propto n^{-1/2}$ (plain),  $v_{\rm A} \propto n^{0}$ (dotted), and $v_{\rm A} \propto n^{0.33}$ (dot-dashed) are plotted for reference. }
\label{fig_v_rho_all}
\end{figure}

The averaged Alfv\'en velocity are also shown in Fig. \ref{fig_v_rho_all} for reference, 
and basically the same behavior is seen. 
The Alfv\'en velocity of the four runs are all supersonic at densities below $10^{10}~\cc$. 
This is probably a necessary criterion for the magnetic field to have an impact on the mass spectra, 
and we basically do not see any effects with even weaker fields. 
In the contrary, the thermal support dominates over the magnetic field at high density, 
and this supports the lack of variation of the peak mass (Paper II).

In the regime where the initial density is low and where thermal support is relatively important with respect to the turbulence. 
Adding a moderate magnetic field (run M04A has the same plasma $\beta$ as run M04) does not change significantly the mass spectrum except for introducing possibly more fluctuations. 
If we regard the magnetic pressure similarly as the thermal pressure with some effective polytropic index for diffuse gas, 
this would result in $\gamma_{0, {\rm eff}} \sim 1$ if $v_{\rm A} \gtrsim c_{\rm s}$ and $\gamma_{0, {\rm eff}} \sim 4/3$ otherwise. 
The second case is not important since the thermal pressure dominates, 
while in the first case the magnetic pressure behaves exactly like the thermal pressure and no significant difference should be expected. 

To summarize shortly, the magnetic field reaches similar (subsonic) values at high densities irrespectively of its initial value, 
and this is why the magnetic field does not play an important role in determining the characteristic mass of the molecular cloud fragmentation. 

\section{Conclusions}
Following the previous study by \citet{Lee18a, Lee18b}, 
we enlarged the parameter space of our simulations by varying the thermal behavior of the diffuse gas and adding the magnetic field. 
We caution that this is a series of numerical experiments, of which the setup does not necessarily correspond to 
usual  astrophysical conditions. 
Nonetheless, such choices of parameters allow to systematically characterize the effect of certain physical mechanisms. 

First of all, we varied the bulk temperature of the isothermal diffuse gas. 
The bulk temperature, together with the mean density that is fixed in our simulations, 
is classically believed to govern the characteristic fragmentation mass of a clump of diffuse gas. 
We have already demonstrated previously that the adiabatic end of the EOS is the determining factor of the peak of the mass spectrum \citep{Lee18b}. 
By varying the low-density end of the EOS, we have confirmed here that it is indeed the high-density end of the EOS that matters, 
not the transition point from the isothermal to adiabatic behavior. 

Secondly, we varied the polytropic index of the low-density gas to replace the isothermal EOS. 
No effect has been observed on the mass spectrum peak as expected. 
On the other hand, there seems to be a slight effect on the high-mass end of the mass spectrum due to a modification of the density PDF. 
We discussed such effect through simple arguments, 
while detailed studies of the density PDF behavior across scales remains to be better understood in future studies. 

Finally, we introduced magnetic field at different strengths into two simulation setups with different initial densities, 
that correspond to models A and C in Paper I. 
We have shown that, despite of adding unphysically strong fields, the mass spectrum of the fragmentation results does not seem to be very much modified.
Only the most magnetized case (M12) shows a more remarkable difference probably due to geometrical effects. 
In all circumstances, during local collapses, the magnetic field fails to reach high values that could provide significant support aside from the thermal pressure, and thus does not affect the characteristic mass of the fragmentation outcome. 
In physical conditions, the magnetic field unlikely plays an important role in terms of determining the mass distribution into fragments.

\appendix
%------------------APPENDIX CALCULATION------------------------
\section{Discussion on effect of $\gamma_0$}\label{append_gam0}
We discuss alternative models that consider the influences of $\gamma_0$. 
Using a value of $\gamma_0$ other than 1 is expected to have effects both on the thermal Jeans mass and the density PDF.
In the main article, we discussed a model where PDF is independent of $\gamma_0$, which is probably an outcome of a global turbulent collapse of the whole cloud (regimes TG and PG). 
Here we consider an alternative model, where $\gamma_0$ does affect the density PDF due to local thermal equilibriums.

With a polytropic index $\gamma_0$, the self-similar equilibrium solution yields a density profile \citep[e.g.][]{Yahil83, Suto88}
\begin{align}\label{eq_rho_r}
\rho \propto r^{-2/ (2-\gamma_0)}.
\end{align}
The volumetric density PDF is obtained by counting the volume at each density given that
\begin{align}\label{eq_PDF_v}
{\cal P}(\rho) d\rho \propto 4\pi r^2 dr.
\end{align}
Dividing each sides of Eq. (\ref{eq_PDF_v}) by $d\rho$, 
this then, after some manipulations, leads to
\begin{align}\label{eq_PDF}
{\cal P}(\rho) \propto  \rho^{(-6+3\gamma_0)/ 2} \propto \rho^{-p},
\end{align}
where $p = 1.2, ~1.5, ~\text{and}~1.95$ for $\gamma_0 = 1.2, ~1, ~\text{and}~ 0.7$, respectively.

The powerlaw indices of the mass spectrum in the PG and PL regimes are obtained by simply inserting Eq. (\ref{eq_PDF}) into Eqs. (\ref{eq_spec_mass_T} and \ref{eq_spec_mass_P}). 
In the TL regime, 
the powerlaw index varies around -7.5 for $\gamma_0$ around 1 and stays negative for $\gamma_0<4/3$. 
In the PL regime, the effect of $\gamma_0$ on the mass-size relation and on the PDF cancel out and the mass spectrum is flat in despite of $\gamma_0$.

The behavior of run T10G07 is compatible with the slope of either the TL regime or the PG regime. 
In the TL case, the difference with respect to the canonical run results from the altered density PDF powerlaw.
In the PG case, this is explained by the movement of the sonic scale.
Since the temperature is fixed at $n_{\rm ad}$,  $\gamma_0<1$ implies that the gas is hotter at low densities and therefore the turbulence becomes sonic at a larger scale. 
The pressure-supported regime is therefore extended to higher masses. 
The value $\gamma_0=0.7$ with respect to 1, with $n_{\rm ad} = 10^{10}$ lowers the Mach number at the initial central density $\sim 6 \times 10^7$ from 22 (see paper I) to $\sim 5$, 
which means that the sonic point is not far from the mean density and the arguments of the TG regime seems to hold.
 
\begin{acknowledgements}
This work was granted access to HPC
   resources of CINES under the allocation x2014047023 made by GENCI (Grand
   Equipement National de Calcul Intensif). 
   This research has received funding from the European Research Council under
   the European Community's Seventh Framework Programme (FP7/2007-2013 Grant  Agreement no. 306483). 
   Y.-N. Lee acknowledges the financial support of the UnivEarthS Labex program at Sorbonne Paris Cit\'e (ANR-10-LABX-0023 and ANR-11-IDEX-0005-02)
\end{acknowledgements}

\bibliographystyle{aa}
\bibliography{ref}

\begin{thebibliography}{30}
\expandafter\ifx\csname natexlab\endcsname\relax\def\natexlab#1{#1}\fi

\bibitem[{{Ballesteros-Paredes} {et~al.}(2015){Ballesteros-Paredes},
  {Hartmann}, {P{\'e}rez-Goytia}, \& {Kuznetsova}}]{BallesterosParedes15}
{Ballesteros-Paredes}, J., {Hartmann}, L.~W., {P{\'e}rez-Goytia}, N., \&
  {Kuznetsova}, A. 2015, \mnras, 452, 566

\bibitem[{{Bastian} {et~al.}(2010){Bastian}, {Covey}, \& {Meyer}}]{Bastian10}
{Bastian}, N., {Covey}, K.~R., \& {Meyer}, M.~R. 2010, \araa, 48, 339

\bibitem[{{Bate} {et~al.}(2003){Bate}, {Bonnell}, \& {Bromm}}]{Bate03}
{Bate}, M.~R., {Bonnell}, I.~A., \& {Bromm}, V. 2003, \mnras, 339, 577

\bibitem[{{Bleuler} \& {Teyssier}(2014)}]{Bleuler14}
{Bleuler}, A. \& {Teyssier}, R. 2014, \mnras, 445, 4015

\bibitem[{{Bonnell} {et~al.}(2003){Bonnell}, {Bate}, \& {Vine}}]{Bonnell03}
{Bonnell}, I.~A., {Bate}, M.~R., \& {Vine}, S.~G. 2003, \mnras, 343, 413

\bibitem[{{Cappellari} {et~al.}(2012){Cappellari}, {McDermid}, {Alatalo},
  {Blitz}, {Bois}, {Bournaud}, {Bureau}, {Crocker}, {Davies}, {Davis}, {de
  Zeeuw}, {Duc}, {Emsellem}, {Khochfar}, {Krajnovi{\'c}}, {Kuntschner},
  {Lablanche}, {Morganti}, {Naab}, {Oosterloo}, {Sarzi}, {Scott}, {Serra},
  {Weijmans}, \& {Young}}]{Cappellari12}
{Cappellari}, M., {McDermid}, R.~M., {Alatalo}, K., {et~al.} 2012, \nat, 484,
  485

\bibitem[{{Chabrier}(2003)}]{Chabrier03}
{Chabrier}, G. 2003, \pasp, 115, 763

\bibitem[{{Girichidis} {et~al.}(2011){Girichidis}, {Federrath}, {Banerjee}, \&
  {Klessen}}]{Girichidis11}
{Girichidis}, P., {Federrath}, C., {Banerjee}, R., \& {Klessen}, R.~S. 2011,
  \mnras, 413, 2741

\bibitem[{{Guszejnov} {et~al.}(2018){Guszejnov}, {Hopkins}, {Grudi{\'c}},
  {Krumholz}, \& {Federrath}}]{Guszejnov18}
{Guszejnov}, D., {Hopkins}, P.~F., {Grudi{\'c}}, M.~Y., {Krumholz}, M.~R., \&
  {Federrath}, C. 2018, \mnras, 480, 182

\bibitem[{{Hennebelle} \& {Chabrier}(2008)}]{HC08}
{Hennebelle}, P. \& {Chabrier}, G. 2008, \apj, 684, 395

\bibitem[{{Hennebelle} \& {Chabrier}(2009)}]{HC09}
{Hennebelle}, P. \& {Chabrier}, G. 2009, \apj, 702, 1428

\bibitem[{{Hopkins}(2012)}]{Hopkins12}
{Hopkins}, P.~F. 2012, \mnras, 423, 2037

\bibitem[{{Hosek} {et~al.}(2018){Hosek}, {Lu}, {Anderson}, {Najarro}, {Ghez},
  {Morris}, {Clarkson}, \& {Albers}}]{Hosek18}
{Hosek}, Jr., M.~W., {Lu}, J.~R., {Anderson}, J., {et~al.} 2018, ArXiv e-prints
  [\eprint[arXiv]{1808.02577}]

\bibitem[{{Inutsuka}(2001)}]{Inutsuka01}
{Inutsuka}, S.-i. 2001, \apjl, 559, L149

\bibitem[{{Kroupa}(2001)}]{Kroupa01}
{Kroupa}, P. 2001, \mnras, 322, 231

\bibitem[{{Krumholz} {et~al.}(2011){Krumholz}, {Klein}, \&
  {McKee}}]{Krumholz11}
{Krumholz}, M.~R., {Klein}, R.~I., \& {McKee}, C.~F. 2011, \apj, 740, 74

\bibitem[{{Lee} \& {Hennebelle}(2018{\natexlab{a}})}]{Lee18a}
{Lee}, Y.-N. \& {Hennebelle}, P. 2018{\natexlab{a}}, \aap, 611, A88

\bibitem[{{Lee} \& {Hennebelle}(2018{\natexlab{b}})}]{Lee18b}
{Lee}, Y.-N. \& {Hennebelle}, P. 2018{\natexlab{b}}, \aap, 611, A89

\bibitem[{{Masunaga} {et~al.}(1998){Masunaga}, {Miyama}, \&
  {Inutsuka}}]{Masunaga98}
{Masunaga}, H., {Miyama}, S.~M., \& {Inutsuka}, S.-i. 1998, \apj, 495, 346

\bibitem[{{Mestel}(1965)}]{Mestel65}
{Mestel}, L. 1965, \qjras, 6, 161

\bibitem[{{Mouschovias}(1991)}]{Mouschovias91}
{Mouschovias}, T.~C. 1991, in NATO Advanced Science Institutes (ASI) Series C,
  Vol. 342, NATO Advanced Science Institutes (ASI) Series C, ed. C.~J. {Lada}
  \& N.~D. {Kylafis}, 61

\bibitem[{{Murray} {et~al.}(2017){Murray}, {Chang}, {Murray}, \&
  {Pittman}}]{Murray17}
{Murray}, D.~W., {Chang}, P., {Murray}, N.~W., \& {Pittman}, J. 2017, \mnras,
  465, 1316

\bibitem[{{Murray} \& {Chang}(2015)}]{Murray15}
{Murray}, N. \& {Chang}, P. 2015, \apj, 804, 44

\bibitem[{{Offner} {et~al.}(2014){Offner}, {Clark}, {Hennebelle}, {Bastian},
  {Bate}, {Hopkins}, {Moraux}, \& {Whitworth}}]{Offner14}
{Offner}, S.~S.~R., {Clark}, P.~C., {Hennebelle}, P., {et~al.} 2014, Protostars
  and Planets VI, 53

\bibitem[{{Offner} {et~al.}(2008){Offner}, {Klein}, \& {McKee}}]{Offner08}
{Offner}, S.~S.~R., {Klein}, R.~I., \& {McKee}, C.~F. 2008, \apj, 686, 1174

\bibitem[{{Padoan} {et~al.}(1997){Padoan}, {Nordlund}, \& {Jones}}]{Padoan97}
{Padoan}, P., {Nordlund}, A., \& {Jones}, B.~J.~T. 1997, \mnras, 288, 145

\bibitem[{{Salpeter}(1955)}]{Salpeter55}
{Salpeter}, E.~E. 1955, \apj, 121, 161

\bibitem[{{Suto} \& {Silk}(1988)}]{Suto88}
{Suto}, Y. \& {Silk}, J. 1988, \apj, 326, 527

\bibitem[{{Vaytet} \& {Haugb{\o}lle}(2017)}]{Vaytet17}
{Vaytet}, N. \& {Haugb{\o}lle}, T. 2017, \aap, 598, A116

\bibitem[{{Yahil}(1983)}]{Yahil83}
{Yahil}, A. 1983, \apj, 265, 1047

\end{thebibliography}

\end{document}